\documentclass[usenatbib,fleqn]{mnras}
 \pdfoutput=1 %for arxiv
\usepackage{algorithm,algorithmic}
\usepackage[T1]{fontenc}
\usepackage{ae,aecompl}

\usepackage{newtxtext,newtxmath}
\usepackage[dvips]{graphicx}
\usepackage{epsfig,color}
\newcommand{\beq}{\begin{equation}}
\newcommand{\eeq}{\end{equation}}
\newcommand{\beqn}{\begin{eqnarray}}
\newcommand{\eeqn}{\end{eqnarray}}

\DeclareMathOperator*{\argmin}{arg\,min}
\def\bmath#1{\mbox{\boldmath$#1$}}

\long\def\symbolfootnote[#1]#2{\begingroup%
\def\thefootnote{\fnsymbol{footnote}}\footnote[#1]{#2}\endgroup}

\usepackage{pstricks,pst-text}
\floatname{algorithm}{Algorithm}

\title[Smart Calibration]{Deep reinforcement learning for smart calibration of radio telescopes}
\author[Yatawatta and Avruch]{Sarod Yatawatta$^{1}$\thanks{E-mail:
yatawatta@astron.nl} and Ian M. Avruch\thanks{E-mail:ian.professional@avruch.net}\\
$^{1}$ASTRON, Oude Hoogeveensedijk 4, 7991 PD Dwingeloo, The Netherlands}

\begin{document}
\date{\today}
\pagerange{\pageref{firstpage}--\pageref{lastpage}} \pubyear{2018}
\maketitle
\label{firstpage}
\begin{abstract}
Modern radio telescopes produce unprecedented amounts of data, which are passed through many
processing pipelines before the delivery of scientific results. Hyperparameters of these pipelines
need to be tuned by hand to produce optimal results. Because many thousands of observations are taken during a lifetime of a telescope and because each observation will have its unique settings, the fine tuning of pipelines is a tedious task. In order to automate this process of hyperparameter selection in data calibration pipelines, we introduce the use of reinforcement learning. {We test two reinforcement learning techniques, twin delayed deep deterministic policy gradient (TD3) and soft actor-critic (SAC),} to train an autonomous agent to perform this fine tuning. For the sake of generalization, we consider the pipeline to be a black-box system where { the summarized state of the performance of the pipeline} is used by the autonomous agent. The autonomous agent trained in this manner is able to determine optimal settings for diverse observations and is therefore able to perform {\em smart} calibration, minimizing the need for human intervention.
\end{abstract}
\section{Introduction}\label{sec:intro}
Data processing pipelines play a crucial role for modern radio
telescopes, where the incoming data flow is reduced in size by many
orders of magnitude before science-ready data are produced. In order
to achieve this, data are passed through many processing steps, for
example, to mitigate radio frequency interference (RFI), to remove
strong confusing sources and to correct for systematic
errors. Calibration is one crucial step in these interferometric
processing pipelines, not only to correct for systematic errors in the
data, but also to subtract strong outlier sources such that weak RFI
mitigation can be carried out. Most modern radio telescopes are
general purpose instruments because they serve diverse science
cases. Therefore, depending on the observational parameters such as
the direction in the sky, observing frequency, observing time (day or
night) etc., the processing pipelines need to be adjusted to realize
the potential quality of the data.

The best parameter settings for each pipeline are mostly determined by
experienced astronomers. This requirement for hand tuning is
problematic for radio telescopes that stream data uninterruptedly. In
this paper, we propose the use of autonomous agents to replace the
human aspect in fine-tuning pipeline settings. We focus on calibration
pipelines, in particular, distributed calibration pipelines where data
at multiple frequencies are calibrated together, in a distributed
computer \citep{DCAL,Brossard2016,DMUX,OLLIER2018,Y2020}. In this
setting, the smoothness of systematic errors across frequency is
exploited to get a better quality solution, as shown by results with
real and simulated data \citep{Patil2017,mertens2020,mevius2021}. { The quality can be further improved by enforcing spatial regularization \citep{Y2020}, which in turn adds more hyperparameters. A comparable situation exists in multispectral image synthesis pipelines as well \citep{Ammanouil2019}, where the determination of the optimal hyperparameters is computationally expensive.}

{ One} key hyperparameter that needs fine-tuning in a distributed
calibration setting is the regularization factor for the smoothness of
solutions across frequency. { Note that spatial smoothness will increase the number of hyperparameters but the methods developed in this paper will be applicable in those cases as well}. It is possible to select the
regularization analytically \citep{EUSIPCO}, however, extending this
to calibration along multiple directions is not optimal. { This is because directions that are spread over a  wide field of view will have highly uncorrelated systematic errors due to the antenna beam and ionosphere, thus requiring the fine-tuning of the regularization per each direction.} In this paper, we
train an autonomous agent to predict the regularization factors for a
distributed, multi-directional calibration pipeline. The calibration
pipeline will perform calibration with a cadence of a few seconds to a
few minutes of data. Therefore, for an observation of long duration
the calibration pipeline needs to be run many times. At the beginning
of every observation the autonomous agent will recommend the
regularization factors to be used in the pipeline based on the pipeline {\em state} and will update the { regularization} factors at the required cadence. { This concept can be extended to train autonomous agents that do more than hyperparamter tuning, for example to optimally allocate compute resources.}

We train the autonomous agent using reinforcement learning (RL) \citep{SuttonBarto}. Reinforcement learning is a branch of machine learning where an agent is trained by repeated interactions with its environment, with proper rewards (or penalties) awarded whenever a correct (or incorrect) decision (or action) is made. In our case, the environment or the observations made by the agent are: i) the input data to the pipeline ii) the input sky model to the pipeline used in calibration iii) the output solutions, and, iv) the output residual data from the pipeline. { In order to minimize the computational cost, we do not feed data directly to the agent. Instead, we only feed the minimum required information of the performance of the pipeline (also called the {\em state}) to the agent.} Because the calibration is done { with a small time-frequency interval,} the amount of data used to determine the state is also small (compared with the full observation). The determination of the reward is done considering two factors. First, the calibration should perform accurately such that the variance of the residual data should be lower than the input data (scaled by the correction applied during calibration). Secondly, we should not overfit the data, which will suppress signals of scientific interest in the residual. { Unlike supervised learning, we do not have access to ground-truth information to determine this overfitting.} Instead, in order to measure the overfitting, we use the influence function \citep{Hampel86,cook1982residuals,Koh17}. { The influence function is useful for assessing a statistical estimator's robustness against errant data.}
  
{
The influence function can be formally described as follows \citep{Hampel86}: Consider a vector statistic ${\bf T}$ which operates on the data vector ${\bf x}$. The cumulative density function (cdf) of the random process with ${\bf x}$ samples is ${\mathcal F}$. The statistic ${\bf T}$ is a composition of operations on the data (a simple example is the mean of the data). The statistic ${\bf T}$ is related to the statistical functional ${\mathcal T}$ as ${\bf T}={\mathcal T}(\widehat{\mathcal F})$ where $\widehat{\mathcal F}$ is the empirical cdf (note that the statistical functional operates on the cdf while the statistic operates on the data). With this setting, we define the influence function $IF_{{\mathcal T},{\mathcal F}}({\bf x})$ as
\beq \nonumber
IF_{{\mathcal T},{\mathcal F}}({\bf x})=\underset{\epsilon \rightarrow 0}{\lim} \frac{{\mathcal T}\left((1-\epsilon) {\mathcal F} + \epsilon \delta_{\bf x} \right) - {\mathcal T}\left({\mathcal F}\right)}{\epsilon}
\eeq
where ${\delta_{\bf x}}$ is a unit point mass probability\footnote{$\delta_{\bf x}$ operates in cdf space, similar to a Heaviside step function.} located at ${\bf x}$. The influence function is a generalization of the derivative for statistical functionals, and can be seen as the sensitivity of ${\mathcal T}$ to a change in the cdf ${\mathcal F}$ due to an additional datum. By studying $IF_{{\mathcal T},{\mathcal F}}({\bf x})$, we can study the behavior of ${\mathcal T}$ to perturbations of data and consequently, we can study the behavior of ${\bf T}$ to perturbations of data.

In this paper, we consider the residual of the data after calibration as the statistic ${\bf T}$. We are interested in studying the effect of model incompleteness on the output residual. One operation in calibration is a minimization (or an ${\argmin(\cdot)}$) and the derivation of the influence function for this case is more involved. We have derived this in our previous work \citep{SAM2018,ST2019}.} We build on this result and determine the reward by considering both the noise reduction and the influence function of calibration. To summarize, we consider both the bias and the variance \citep{neco1992} introduced by the pipeline for the determination of the state as well as the reward. The agent will perform an action, that is to recommend the optimal regularization factors to use in the calibration pipeline. { Compared to conventional methods of hyperparameter fine-tuning \citep{bergstra12a,Bates2021}, the agent will make a decision much faster, but on the other hand, the training of the agent will incur additional initial cost.}

The advent of deep neural networks (deep learning) \citep{Lecun2015} combined with better training algorithms have enabled RL agents to perform superhuman tasks. Early RL breakthroughs were made in discrete action spaces such as in computer games \citep{Atari}. The regularization factors used in calibration are in a continuous action space, and therefore, in this paper we use RL techniques suitable for continuous action spaces, namely, deep deterministic policy gradient (DDPG) \citep{DDPG} and its improvements, TD3 \citep{TD3} and SAC \citep{SAC}. In this paper, we only consider calibration pipelines that perform distributed, direction dependent calibration, but the same RL technique can be used in other data processing pipelines in radio astronomy and beyond.

The rest of the paper is organized is as follows. In section \ref{sec:model}, we describe the distributed, direction dependent calibration pipeline. In section \ref{sec:RL}, we introduce our RL agent and its environment. In section \ref{sec:simul}, we provide results based on simulated data to show the efficacy of the trained RL agent. Finally, we draw our conclusions in section \ref{sec:conc}.

{\em Notation}: Lower case bold letters refer to column vectors (e.g. ${\bf y}$). Upper case bold letters refer to matrices (e.g. ${\bf C}$). Unless otherwise stated, all parameters are complex numbers. The set of complex numbers is given as ${\mathbb C}$ and the set of real numbers as  ${\mathbb R}$. The matrix inverse, pseudo-inverse, transpose, Hermitian transpose, and conjugation are referred to as $(\cdot)^{-1}$, $(\cdot)^{\dagger}$, $(\cdot)^{T}$, $(\cdot)^{H}$, $(\cdot)^{\star}$, respectively. The matrix Kronecker product is given by $\otimes$. The vectorized representation of a matrix is given by $\mathrm{vec}(\cdot)$. The identity matrix of size $N$ is given by ${\bf I}_N$. All logarithms are to the base $e$, unless stated otherwise. The Frobenius norm is given by $\|\cdot \|$ and the L1 norm is given by $\|\cdot \|_1$.

\section{Distributed direction dependent calibration pipeline}\label{sec:model}
{
In this section, we first give a brief overview of the distributed, direction dependent calibration pipeline that is used by the RL agent for hyperparemeter tuning. Thereafter, we give an overview of performance analysis of the calibration pipeline using influence functions. 

\subsection{Distributed calibration}
}
We refer the reader to \cite{DCAL,Y2020} for an in-depth overview of distributed calibration. We outline the basic concepts here.

The signal { (in full polarization)} received by an interferometer is given by \citep{HBS}
\beq \label{V}
{\bf V}_{pqf}=\sum_{k=1}^{K} {\bf J}_{kpf} {\bf C}_{kpqf} {\bf J}_{kqf}^H + {\bf N}_{pqf}
\eeq
where we have signals from $K$ discrete sources in the sky being received. All items in (\ref{V}), i.e., ${\bf V}_{pqf},{\bf C}_{kpqf}, {\bf J}_{kpf}, {\bf J}_{kqf}, {\bf N}_{pqf} \in {\mathbb C}^{2\times 2}$ are implicitly time varying and subscripts $p,q$ correspond to stations $p$ and $q$, $f$ correspond to the receiving frequency, and $k$ correspond to the source index. The coherency of the $k$-th source at baseline $p$-$q$ is given by ${\bf C}_{kpqf}$ and for a known source, this can be pre-computed \citep{TMS}.  The noise ${\bf N}_{pqf}$ consists of the actual thermal noise as well as signals from sources in the sky that are not part of the model. The systematic errors due to the propagation path (ionosphere), receiver beam, and receiver electronics are represented by ${\bf J}_{kpf}$ and ${\bf J}_{kqf}$. For each time and frequency, with $N$ stations, we have $N(N-1)/2$ pairs of $p$-$q$.

Calibration is the estimation of ${\bf J}_{kpf}$ for all $k$,$p$ and $f$, using data within a small time interval { (say $\Delta_t$)}, within which it is assumed that ${\bf J}_{kpf}$ remains constant. An important distinction is that we only calibrate along $K$ directions, but in real observations, there { are} signals from faint sources (including the Galaxy) that are not being modeled. We are not presenting a method of calibration in this paper, and therefore the actual method of calibration is not relevant. However, for the sake of simplicity, we follow a straightforward description. What we intend to achieve is to relate the hyperparameters used in the pipeline to the performance of calibration. Therefore, we represent ${\bf J}_{kpf}$ { ($\in {\mathbb C}^{2\times 2}$)} as $8$ real parameters (for frequency $f$ and direction $k$), and we have $8\times N\times K$ real parameters in total for frequency $f$, per direction, namely ${\bmath \theta}_{kf}$.

We represent { the matrix equation} (\ref{V}) in vector form as
\beq \label{vecV}
{\bf v}_{pqf}=\sum_{k=1}^K {\bf s}_{kpqf}({\bmath \theta}_{kf}) +{\bf n}_{pqf}
\eeq
where ${\bf s}_{kpqf}({\bmath \theta}_{kf})=\left({\bf J}_{kqf}^{\star}\otimes{\bf J}_{kpf}\right) vec({\bf C}_{kpqf})$, ${\bf v}_{pqf}=vec({\bf V}_{pqf})$, and ${\bf n}_{pqf}=vec({\bf N}_{pqf})$. 

We stack up the data and model vectors of (\ref{vecV}) for the { $\Delta_t$} time slots within which a single solution is obtained as
\beqn \label{modvec}
{\bf x}_f&&=[\mathrm{real}({\bf v}_{12}^T),\mathrm{imag}({\bf v}_{12}^T),\mathrm{real}({\bf v}_{13}^T),\ldots]^T\\\nonumber
 {\bf s}_{kf}({\bmath \theta}_{kf})&&=[\mathrm{real}\left({\bf s}_{k12f}({\bmath \theta})^T\right),\mathrm{imag}\left({\bf s}_{k12f}({\bmath \theta})^T\right),\mathrm{real}\left({\bf s}_{k13f}({\bmath \theta})^T\right),\\\nonumber
 &&\ldots]^T
\eeqn
which are vectors of size { $8\times N(N-1)/2 \times \Delta_t$} and we have 
\beq \label{xvec}
{\bf x}_f = \sum_{k=1}^K {\bf s}_{kf}({\bmath \theta}_{kf}) + {\bf n}_{f}={\bf s}_{f}({\bmath \theta}_{f}) + {\bf n}_{f}
\eeq
as our final measurement equation { with real valued vectors} ${\bmath \theta}_{kf} \in {\mathbb R}^{8N}$, ${\bmath \theta}_{f} \in {\mathbb R}^{8KN}$.

In order to estimate ${\bmath \theta}_{kf}$  from (\ref{xvec}), we define a loss function
\beq \label{loss}
g_{kf}({\bmath \theta}_{kf})=loss({\bf x}_f,{\bmath \theta}_{kf})
\eeq
depending on the noise model, for example, mean squared error loss, $\|{\bf x}_f - {\bf s}_{f}({\bmath \theta}_{f}) \|^2$. We calibrate data at many frequencies and we exploit the smoothness of ${\bmath \theta}_{kf}$ with $f$ and in order to make the solutions smooth. We introduce the constraint
\beq \label{constraint}
{\bmath \theta}_{kf} = {\bf Z}_k {\bf b}_f
\eeq
where ${\bf b}_f$ ($\in {\mathbb R}^{P}$) is a polynomial basis { vector in frequency  ($P$ terms) evaluated at $f$ and ${\bf Z}_k$ ($\in {\mathbb R}^{8N \times P}$) is a matrix that shares information about the contribution of each basis function over all frequencies. In addition, we can introduce constraints for spatial smoothness (using $B$ bases in space) as 
\beq \label{spatial}
{\bf Z}_k = \overline{\bf Z}_k = {\bf Z} {\bf B}_k
\eeq
where ${\bf B}_k$ is a basis matrix ($\in {\mathbb R}^{B \times P}$) evaluated along the direction $k$ and ${\bf Z}$ ($\in {\mathbb R}^{8N \times B}$) is a matrix that shares information common to all $K$ directions in the sky.
}

{
We introduce Lagrange multipliers ${\bf w}_{kf}$ ($\in {\mathbb R}^{8N}$) for the constraint (\ref{constraint}) and ${\bf W}_{k}$ ($\in {\mathbb R}^{8N \times P}$) for the constraint (\ref{spatial}). We use the augmented Lagrangian method (or method of multipliers) \citep{Hestenes69,Powell69} to solve this. We write the augmented Lagrangian as
\beqn \label{aug}
\lefteqn{
l({\bmath \theta}_{kf},{\bf w}_{kf},{\bf W}_k,{\bf Z}_k,{\bf Z}:\forall k,f)=}\\\nonumber
&&\sum_{kf} g_{kf}({\bmath \theta}_{kf}) + {\bf w}_{kf}^T({\bmath \theta}_{kf} - {\bf Z}_k {\bf b}_f)+\frac{\rho_k}{2}\|{\bmath \theta}_{kf} - {\bf Z}_k {\bf b}_f\|^2\\\nonumber
&&+\sum_k {\bf W}_k^T({\bf Z}_k-\overline{\bf Z}_k)+\frac{\alpha_k}{2} \|{\bf Z}_k-\overline{\bf Z}_k \|^2
\eeqn
which we minimize to find a smooth solution. As described in \citep{Y2020}, we use a combination of alternating direction method of multipliers \citep{boyd2011} and (weighted) federated averaging \citep{McMahan2016} to solve this.
}

{
The hyperparameters in (\ref{aug}) are the regularization factors $\alpha_k,\rho_k$ ($\in {\mathbb R}^+$), where we have two regularization factors per each direction $k$. However, for the remainder of the paper, for simplicity, we only consider the constraint (\ref{constraint}) and ignore the constraint (\ref{spatial}).} Therefore, we need to select $K$ positive, real valued hyperparameters to get the optimal solution to (\ref{aug}). We can consider the $K$ directional calibration as $K$ separate sub problems and find analytical solutions to optimal $\rho_k$ as done in \cite{EUSIPCO}. { However, by considering each direction separately, we ignore the inter-dependencies between directions, thereby making our optimal $\rho_k$ only an approximation}. { If we consider the constraint (\ref{spatial}) as well, the complexity of selecting the optimal $\alpha_k,\rho_k$ increases.} The alternative is to use handcrafted methods such as grid search but this is infeasible, as every observation will have unique requirements for grid search hyperparameters. In section \ref{sec:RL}, we will present the use of RL to automatically find the best values for $\rho_k$ { considering all $K$ directions together (which can be extended to find $\alpha_k$ as well)}.

{
\subsection{Performance analysis using the influence function}
}
The RL agent interacts with the pipeline (the pipeline is solving (\ref{aug}), see Fig. \ref{agent_pipeline}) to recommend the best values for $\rho_k$. In order to do this, the RL agent needs to have some information about the environment (i.e., the performance of the pipeline). { The performance is measured using the influence function (see section \ref{sec:intro}) of the output statistic, i.e., the residual.} After each calibration, we calculate the residual using (\ref{V}) as
\beq \label{res}
{\bf R}_{pqf}={\bf V}_{pqf}-\sum_{k=1}^{K} \widehat{{\bf J}}_{kpf} {\bf C}_{kpqf} \widehat{{\bf J}}_{kqf}^H
\eeq
where $\widehat{{\bf J}}_{kpf} \forall k,p,f$ are calculated using the solutions obtained for ${\bmath \theta}_{kf}$.

{ In order to analyze the influence function of the residual, we reformulate our data model using real vectors.} We can rewrite (\ref{xvec}) in a simplified form (dropping the subscripts $k,f$) as 
\beq \label{smodel}
{\bf x}={\bf s}({\bmath \theta}) + {\bf n}
\eeq
where ${\bf x}$ ($\in \mathbb{R}^D$) is the data ($D=8 \times N(N-1)/2 \times \Delta_t$), ${\bf s}({\bmath \theta})$ is our model and ${\bf n}$ is the noise (noise as defined in (\ref{V}) which includes unmodeled sources in the field of view). We find a solution to ${\bmath \theta}$ by minimizing a loss $g({\bmath \theta})$, say $\widehat{{\bmath \theta}} = {\argmin}_{\bmath \theta}\  g({\bmath \theta})$. Thence, we calculate the residual as
\beq \label{resvec}
{\bf y}={\bf x}-{\bf s}(\widehat{{\bmath \theta}}).
\eeq
{ To recapitulate, ${\bf x}$ in (\ref{resvec}) is the vectorized version of ${\bf V}_{pqf}$ in (\ref{res}) and ${\bf y}$ is the vectorized version of ${\bf R}_{pqf}$, for all $p,q$ within $\Delta_t$ (real and imaginary parts separate). Referring back to our definition of influence function in section \ref{sec:intro}, we consider the residual ${\bf y}$ to be a statistic that is derived from the data ${\bf x}$.} We have shown \citep{ST2019} that the probability density functions (pdf) of ${\bf x}$ and ${\bf y}$, i.e., $p_X({\bf x})$,  $p_Y({\bf y})$ are related as
\beq \label{pdf}
p_X({\bf x}) = |\mathcal{J}| p_Y({\bf y})
\eeq
where $|\mathcal{J}|$ is the Jacobian determinant of the mapping from ${\bf x}$ to ${\bf y}$. { We can use (\ref{pdf}) to find the mapping between the cumulative density functions and relate it to the formal definition of $IF_{{\mathcal T},{\mathcal F}}({\bf x})$ in section \ref{sec:intro}, but for our purpose, working with the pdf relation (\ref{pdf}) is sufficient.} Using (\ref{resvec}), we can show that 
\beq
|\mathcal{J}|=\exp \left( \sum_{i=1}^D \log(1 + \lambda_i(\mathcal{A})) \right)
\eeq
where 
\beq \label{calA}
\mathcal{A} = \frac{\partial {\bf s}({\bmath \theta})}{\partial {\bmath \theta}^T} \left( \frac{\partial^2 g({\bmath \theta})}{\partial {\bmath \theta} \partial {\bmath \theta}^T} \right)^{-1} \frac{\partial^2 g({\bmath \theta})}{\partial {\bmath \theta} \partial {\bf x}^T}
\eeq
evaluated at ${\bmath \theta}=\widehat{{\bmath \theta}}$. Ideally, we should have $|\mathcal{J}|=1$ and therefore all eigenvalues of $\mathcal{A}$, $\lambda_i(\mathcal{A})=0$. But in practice, due to model error and degeneracies in the model, we have some non-zero values for $\lambda_i(\mathcal{A})$.

\begin{figure*}
\begin{minipage}{0.98\linewidth}
\begin{center}
\begin{minipage}{0.48\linewidth}
\centering
 \centerline{\epsfig{figure=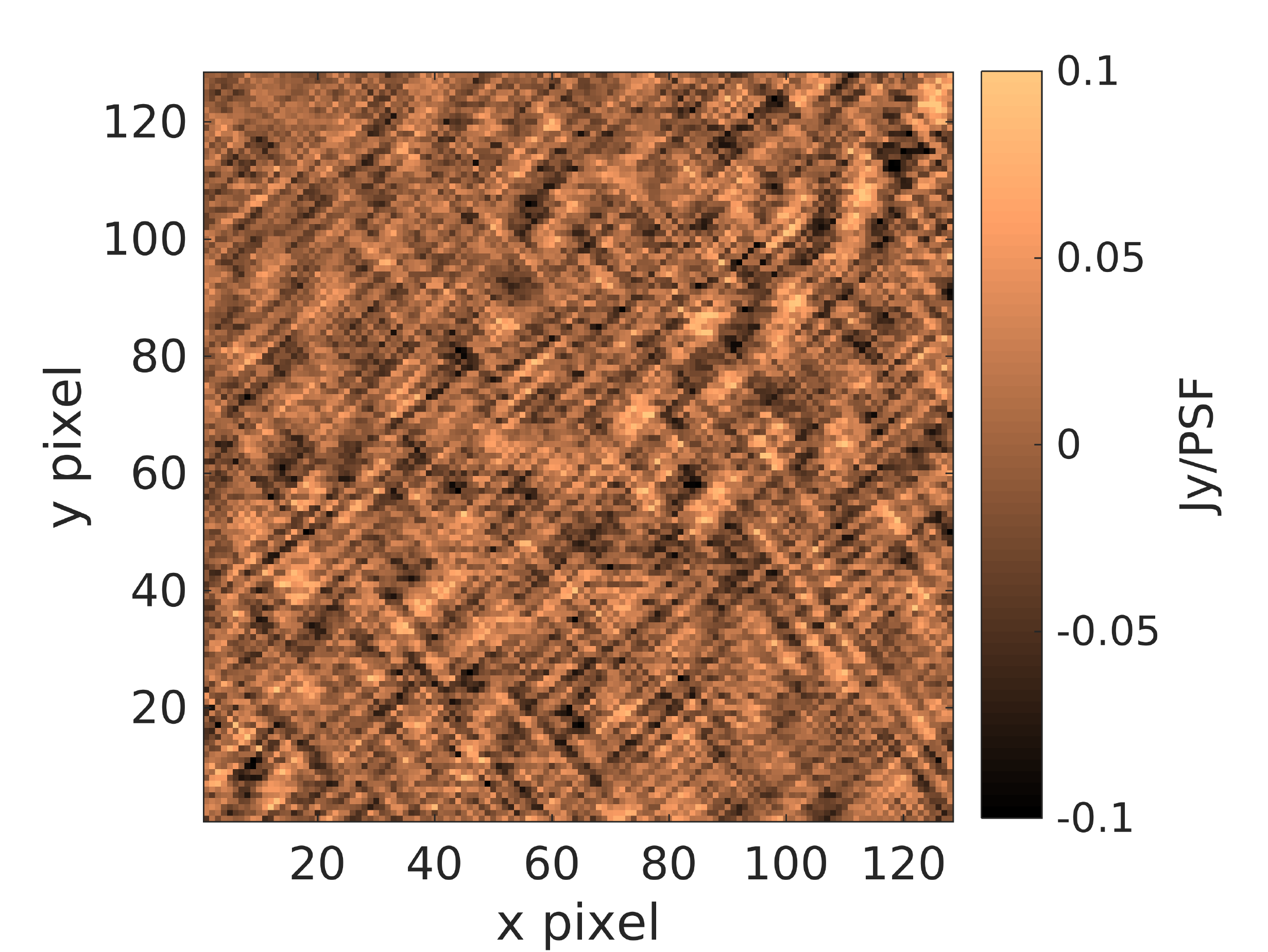,width=8.0cm}}
\vspace{0.5cm} \centerline{(a)}\smallskip
\end{minipage}
\begin{minipage}{0.48\linewidth}
\centering
 \centerline{\epsfig{figure=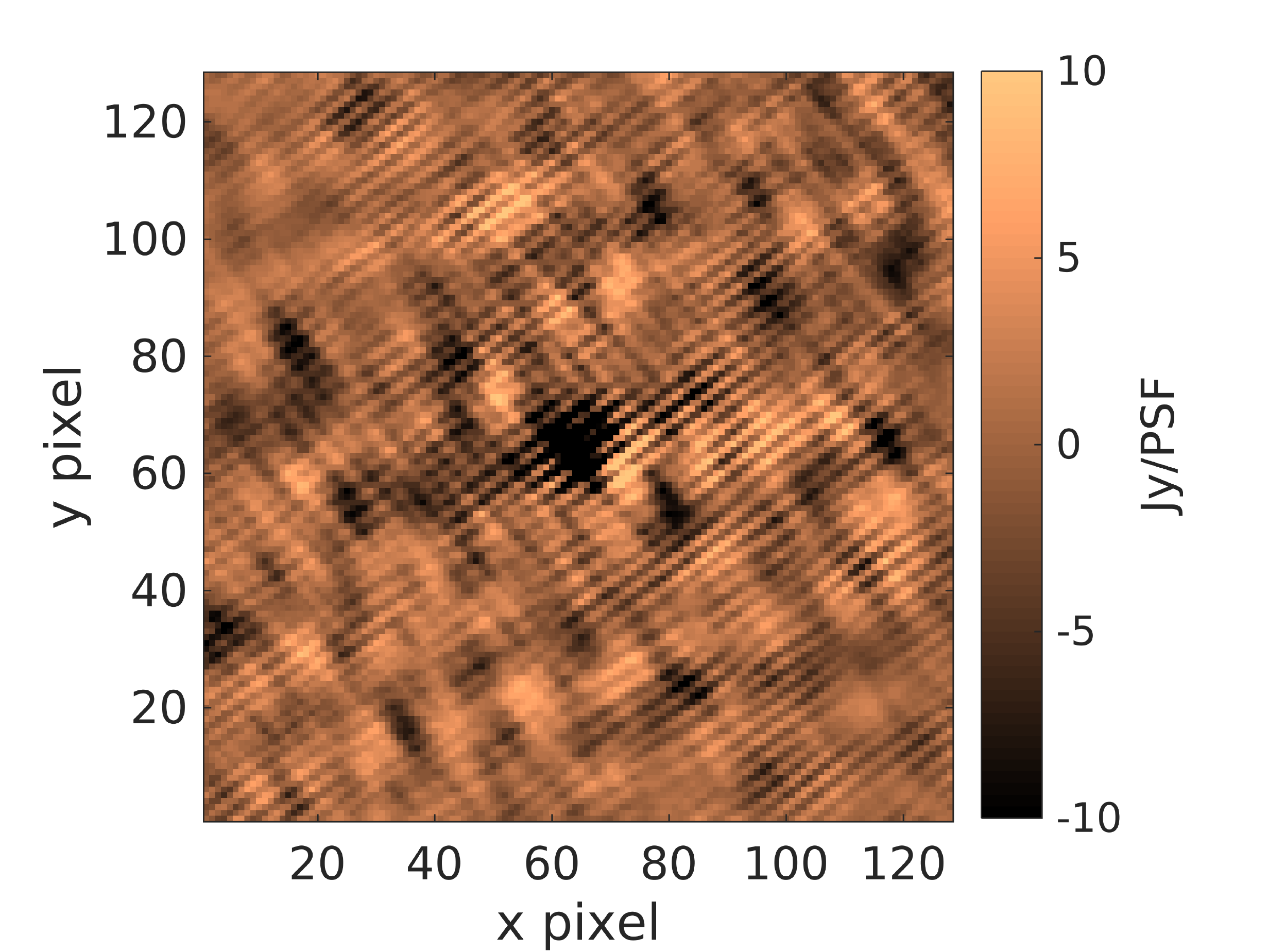,width=8.0cm}}
\vspace{0.5cm} \centerline{(b)}\smallskip
\end{minipage}\\
\begin{minipage}{0.48\linewidth}
\centering
 \centerline{\epsfig{figure=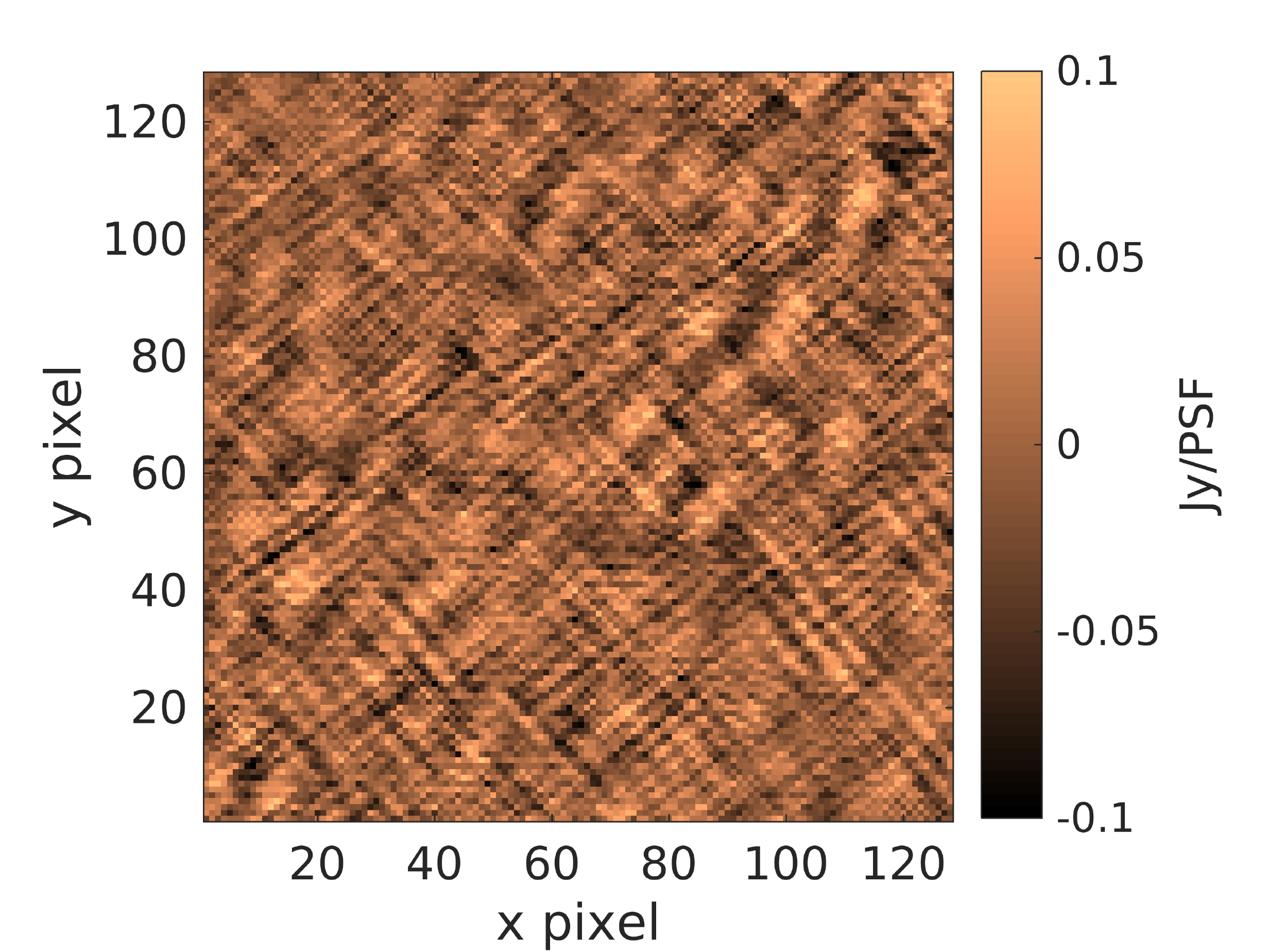,width=8.0cm}}
\vspace{0.5cm} \centerline{(c)}\smallskip
\end{minipage}
\begin{minipage}{0.48\linewidth}
\centering
 \centerline{\epsfig{figure=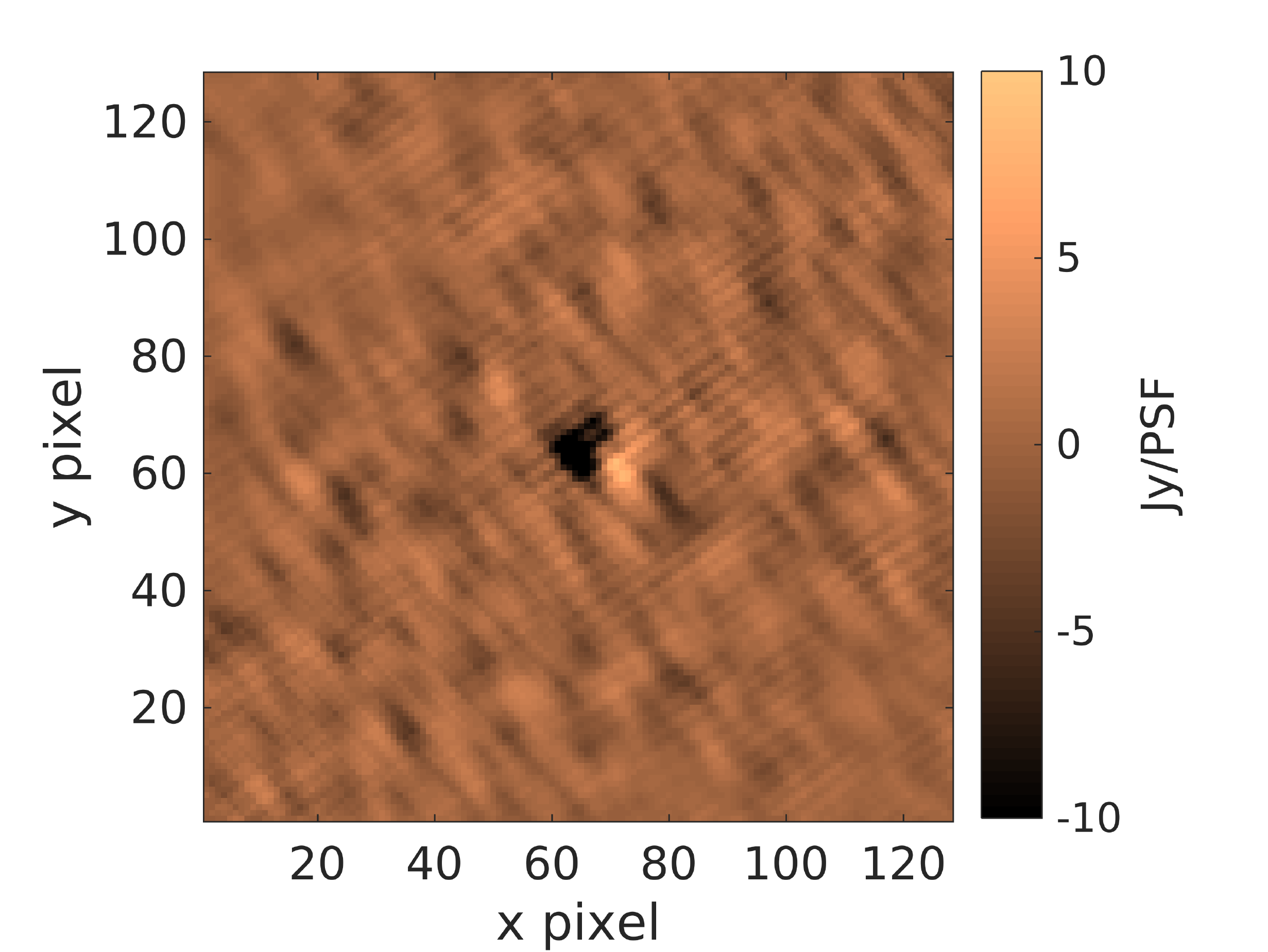,width=8.0cm}}
\vspace{0.5cm} \centerline{(d)}\smallskip
\end{minipage}
\end{center}
\caption{
{ The residual images (left panels) and corresponding influence maps (right panels) for two calibration scenarios, using 1 min of data.  The bottom row (panels (c) \& (d)) is the optimally-regularized solution. With this small amount of data the influence map change is dramatic, while the residual map change is
   subtle. We use the influence map as a sensitive indicator of solution quality to train our agent by reinforcement learning.}
\label{influence_maps}}
\end{minipage}
\end{figure*}

{ 
Let us denote the $n$-th element of ${\bf x}$ in (\ref{resvec}) as $x_n$. As ${\bf x}$ is simply a reorganization of ${\bf V}_{p^\prime q^\prime f}$ in (\ref{res}), given any value of $n$ ($\in[0,D-1]$), we can find the corresponding baseline $p^\prime-q^\prime$ as well as the corresponding element in the matrix ${\bf V}_{p^\prime q^\prime f}$ and whether it represents the real or imaginary part of that element. Let us use $l$ ($\in[0,7]$, $4$ complex correlation values) to denote this index in the matrix ${\bf V}_{p^\prime q^\prime f}$. Note that for the same baseline $p^\prime-q^\prime$ and $l$, we will have $\Delta_t$ values in the vector ${\bf x}$ because we use multiple time ordered samples from the same baseline. Nonetheless, given $n$, we can find the corresponding $p^\prime-q^\prime$ and $l$ and vice versa. Therefore, we use the notation $x_n$ and $x_{p^\prime q^\prime l}$ to denote the same element in the vector ${\bf x}$. A similar relation exists between the $n$-th element of ${\bf y}$ in (\ref{resvec}) and the corresponding element of ${\bf R}_{pqf}$ in (\ref{res}).

Using (\ref{calA}), and using the dual notation described above, we can calculate $\frac{\partial {\bf y}}{\partial x_{p^\prime q^\prime l f}}$. Reordering the matching $8$ values of $\frac{\partial {\bf y}}{\partial x_{p^\prime q^\prime l f}}$ as a matrix $\in {\mathbb C}^{2\times 2}$, we can also find $\frac{\partial {\bf R}_{pqf}}{\partial x_{p^\prime q^\prime l f}}$. Finally, we find $E\{ \frac{\partial {\bf R}_{pqf}}{\partial x_{p^\prime q^\prime l f}}\}$ where the expectation $E\{\cdot\}$ is taken over $p^\prime$, $q^\prime$ (excluding the tuple $p,q$) and $l$ ($\in[0,7]$). Note that $E\{ \frac{\partial {\bf R}_{pqf}}{\partial x_{p^\prime q^\prime l f}}\}$ has the same size as ${\bf R}_{pqf}$. Therefore, we can replace the residual ${\bf R}_{pqf}$ with $E\{ \frac{\partial {\bf R}_{pqf}}{\partial x_{p^\prime q^\prime l f}}\}$ when we produce the output and feed this to imaging, just like we make images of the residual.  We call the image made in the usual radio astronomical image synthesis manner, replacing the residual ${\bf R}_{pqf}$ with $D \times E\{ \frac{\partial {\bf R}_{pqf}}{\partial x_{p^\prime q^\prime l f}}\}$, as the {\em influence map}, which we denote by $\mathcal{I}({\bmath \theta})$.

We illustrate the relationship between the (Stokes I) images made using the residual and its influence function in Fig. \ref{influence_maps}. We have simulated an observation (see section \ref{ssec:simul_calib}) with $K=10$ directions in the sky being calibrated. The sky model has errors due to weak sources that are not being part of calibration. All images are centred at the phase centre, with size $128\times 128$ pixels (pixel size $20^{\prime\prime} \times 20^{\prime\prime}$), and are made using uniform imaging weights. An important fact is that the images are made using only 1 min of data (which is far less than the full observation).  The left column (panels a and c) in Fig. \ref{influence_maps} shows the residual while the right column (panels b and d) shows the influence map $\mathcal{I}({\bmath \theta})$. The top row is with low regularization (about $1/10$ of the optimal value) and the bottom row is with approximately optimal regularization. Images of the residual show hardly any difference as seen in Fig. \ref{influence_maps} (a) and (c). This is expected because we only use 1 min of data and the subtle differences due to imperfect calibration will only become apparent after processing more data (at least several hours). In contrast, the influence maps in Fig. \ref{influence_maps} (b) and (d) clearly show a difference. Therefore, we conclude that by using $\mathcal{I}({\bmath \theta})$, with minimal amount of data, we are able to measure the performance of calibration. In section \ref{sec:RL}, we will discuss how to train deep neural networks to detect changes in the influence maps and evaluate the performance of calibration.
}

{
We relate the influence map to the bias and variance of calibration residual as follows. The bias is primarily caused by the errors in the sky model (difference between the true sky and the sky model used in calibration). If we consider the solutions to calibration, i.e. $\widehat{{\bmath \theta}}$, with low regularization we will have low bias and high variance of the solutions (overfitting compared to the ground truth) and with high regularization, we will have high bias and low variance of the solutions (underfitting). We can reduce the bias by increasing the model complexity or by decreasing the regularization and conversely, we can reduce the variance by increasing the regularization. From Fig. \ref{influence_maps} we see that low regularization leads to high influence and therefore the variance is directly related to the influence. We relate this result to the residual data ${\bf y}$ in (\ref{resvec}) as well. With high variance of $\widehat{{\bmath \theta}}$, the coherent information hidden in ${\bf y}$ are affected with high variance. In other words, there is a loss of coherence and the coherent weak signals in ${\bf y}$ get suppressed. The fine tuning of $\rho_k$ will give us the optimal tradeoff between bias and the variance of the residual data. An example of this behavior is shown in Fig. \ref{reward}.  We achieve this fine-tuning using RL as described in section \ref{sec:RL}.
}

\section{Reinforcement learning}\label{sec:RL}
\begin{figure}
\begin{minipage}{0.98\linewidth}
\begin{center}
\centering
\centerline{\epsfig{figure=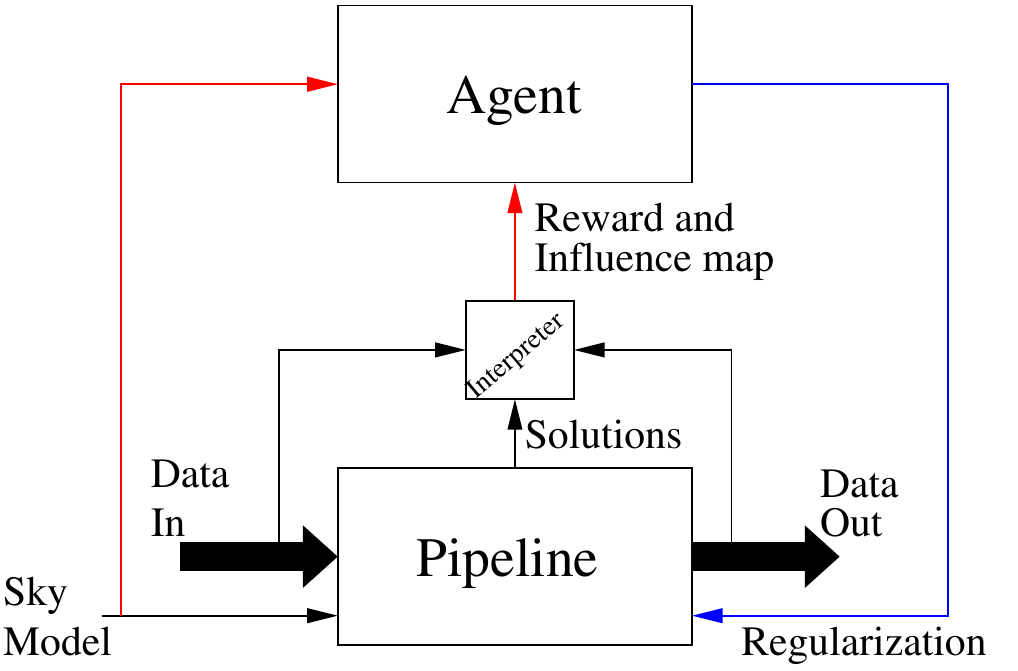,width=8.0cm}}
\vspace{0.1in}
\end{center}
\end{minipage}
\caption{The interactions between the RL agent and its environment, i.e., the pipeline. The pipeline processes data in a stream and only a summarized state of the performance of the pipeline, calculated by the interpreter, is sent to the agent.\label{agent_pipeline}}
\end{figure}

In this section, we describe the basic concepts of reinforcement learning, especially tailored to our problem and thereafter we go into specific details about our RL agent. An RL agent interacts with its environment via some observations and performs an action depending on these observations and the current state of the agent. The objective of the agent is to maximize its reward depending on the action it performs. As shown in Fig. \ref{agent_pipeline}, our agent interacts with our calibration pipeline, which is the environment. We describe these interactions as follows:
\begin{itemize}
\item At an instance indexed by $t$, the input to the agent is given by state $s_t$ (a real valued tensor). This is a summary of the observation that we deem sufficient for the agent to evaluate its action. In our case, we give the sky model used in calibration as part of $s_t$. In addition, we create an influence map $I({\bmath \theta})$, that summarizes the performance of calibration as part of $s_t$. Note that we do not feed the data or the metadata (such as the $uv$ coverage) into the agent. {  This way, our agent will be able to interact with pipelines operating at variable data rates and with variable solution intervals in calibration. The amount of data and the metadata are much larger than the size of $s_t$. Hence, the computational burden of the agent will also be reduced.}
\item Depending on $s_t$ and its internal parameter values, the agent performs { (or recommends)} an action $a_t$ (a real valued vector). In our case, the action is the specification of the regularization factors $\rho_k$ ($K$ values) which is fed back to the pipeline.
\item We determine the reward $r_t$ (a real valued scalar) to the agent using two criteria. First, we compare the noise of the input data $\sigma_0$ and the output residual $\sigma_1$ of the pipeline (by making images of the field being observed { as in Fig. \ref{influence_maps} left column or by time differencing the data}) and find the ratio $\frac{\sigma_0}{\sigma_1}$. We get a higher reward as this ratio grows. Second, we look at the influence map and find its standard deviation, whose inverse we add to the reward. If we have high influence, we will get a smaller reward. With both these components constituting the reward, we aim to minimize both the bias and the variance of the pipeline.
\end{itemize}
Note that unlike in most RL scenarios, we do not have a terminal state (such as in a computer game where a player wins or loses). { We terminate when we reach $L$ iterations of refinement on $\rho_k$. The general interactions between the RL agent and the environment are given in Algorithm \ref{algRL} and we make some remarks about Algorithm \ref{algRL} here:}

{
\begin{algorithm}
\caption{Training the RL agent}
\label{algRL}
\begin{algorithmic}[1]
\REQUIRE Number of episodes $E$, Number of loop iterations $L$
\STATE Initialize Environment and Agent
\STATE Setup empty (or reload saved) ReplayBuffer
\FOR{$e=1,\ldots,E$}
\STATE Environment: simulate new (random) observation
\STATE Agent: $\leftarrow$ initial state $s_1$
\FOR{$t=1,\ldots,L$}
 \STATE \COMMENT{Agent starts interactions}
 \STATE $a_t \leftarrow$ Agent suggests action based on $s_t$
 \STATE Environmet: take action $a_t$, determine reward $r_t$ and next state $s_{t+1}$
 \STATE ReplayBuffer: store $(s_t,a_t,r_t,s_{t+1})$
 \STATE Agent: sample a mini-batch from ReplayBuffer and learn
\ENDFOR
\ENDFOR
\end{algorithmic}
\end{algorithm}
}

{
\begin{itemize}
\item The interactions shown in Algorithm \ref{algRL} are for the learning or training phase of the RL agent. In this training phase, we simulate realistic observations (line 4) which we describe in section \ref{ssec:simul_calib}. The number of simulations (or episodes) is $E$.
\item After the agent has been sufficiently trained, we can use it in processing real observations. In other words, we will have only one episode, $E=1$ and line 4 of Algorithm \ref{algRL} is replaced by the data from the observation. The learning step in line 11 is not necessary.
\end{itemize}

The key step in Algorithm \ref{algRL} is the learning step (line 11).  Because our actions are continuous numbers, the RL agent works in continuous action spaces. There are several RL algorithms that we can use for this step, namely TD3 \citep{TD3} and SAC \citep{SAC}. We give a brief a overview of the inner workings of our RL agent, and for a more thorough overview, we refer the reader to \citep{Silver2014,DDPG,TD3,SAC}. The major components of the agent are:
\begin{itemize} 
\item Actor: a deterministic actor with policy $\pi(s)$ will produce action $a$ given the state $s$, $\pi(s)\rightarrow a$. In contrast, a stochastic actor will predict the conditional probability of an action, $\pi(a|s)$ given the state $s$. We can sample from this conditional distribution to get the action, $a\sim \pi(a|s)$.
\item Critic: given the state $s$ and taking the action $a$, the critic will evaluate the expected return $q$ (a scalar), $Q(s,a) \rightarrow q$. The mapping $Q(\cdot,\cdot)$ is also called the state-action value function.
\item Value: the value function $V(\cdot)$ will produce a scalar output given the state $s$, $V(s) \rightarrow v$. The value $v$ gives a measure of the importance of the state $s$. In contrast to $Q(s,a)$, $V(s)$ is independent of the action taken at state $s$ (but both follow the policy $\pi(\cdot)$).
\item Replay buffer: the replay buffer stores the tuple (state $s$, action $a$, reward $r$, next state $s^\prime$) for each iteration $t$ in Algorithm \ref{algRL}. This acts as a collection of past experience that helps the learning process.
\end{itemize}
}
Note that neither the actor, the critic nor the value function are dependent on past states, but only the current state and a measure of the state's favorability for future rewards, which is a characteristic of a Markov decision process  \citep{SuttonBarto}. Looking into the future, the expected cumulative reward can be written as the Bellman equation
\beq\label{bellman}
Q(s,a)= r + \gamma \underset{a^\prime=\pi(s^\prime)}{\mathrm max} Q(s^\prime,a^\prime)
\eeq
where $s^\prime,a^\prime$ is the (optimal) state action pair for the next step and  $\gamma\approx 1$ is the discount factor for future uncertainty.

\begin{table}
\centering
\begin{tabular}{c|l l|}
 & TD3 & SAC \\
\hline
Critic & $Q_{\xi_1}(s,a)$ and $Q_{\xi_2}(s,a)$ & $Q_{\xi_1}(s,a)$ and $Q_{\xi_2}(s,a)$ \\
& target & \\
& $Q_{\overline{\xi}_1}(s,a)$ and $Q_{\overline{\xi}_2}(s,a)$ & \\
Value & - & $V_{\zeta}(s)$\\
& & target $V_{\overline{\zeta}}(s)$\\
Actor & deterministic actor & stochastic actor \\
& $\pi_{\phi}(s)$ & action sampled from \\ 
 & target $\pi_{\overline{\phi}}(s)$ & conditional $\pi_{\phi}(a|s)$\\
Reward & $r(s,a)$ & $r(s,a)+\alpha \mathcal{H}(\pi(a|s))$\\
& & $\mathcal{H}(\cdot)$ is the entropy\\
& &  reward is scaled by $1/\alpha$ \\
\end{tabular}
\caption{Comparison of the key features of TD3 and SAC.} \label{sac_td3}
\end{table}

We model the actor, the critic and the value as deep neural networks with trainable parameters $\phi$ for the actor, $\xi$ for the critic, and $\zeta$ for the value function. The use of these networks changes depending on the RL algorithm (TD3 or SAC). We list the main differences between TD3 and SAC in Table \ref{sac_td3}.  In TD3, we train the actor and the critic as follows. Given the current state $s$, the actor tries to maximize $J(\phi)=Q(s,a)|_{a=\pi(s)}$ by choice of action. We perform gradient ascent on $J(\phi)$, with the gradient (using the chain rule) given as
\beq \label{gradA}
\nabla_{\phi} J(\phi)=\nabla_a Q(s,a)|_{a=\pi(s)} \times \nabla_\phi \pi(s).
\eeq
The critic is trained by minimizing the error between the current state-action value $Q(s,a)$ and the expected cumulative reward under optimal settings (\ref{bellman})
\beq \label{lossC}
J(\xi)=\|Q_\xi(s,a)-(r + \gamma \underset{a^\prime=\pi(s^\prime)}Q(s^\prime,a^\prime))\|^2.
\eeq

In SAC, we train the value function by minimizing the cost 
\beq \label{lossV}
J(\zeta)=\|V_{\zeta}(s)-Q_\xi(s,a)+\log \pi_{\phi}(a|s)\|^2
\eeq
and thereafter, we train the state-action value function by minimizing
\beq \label{lossCSAC}
J(\xi)=\|Q_\xi(s,a)-(r + \gamma V_{\zeta}(s^\prime) )\|^2.
\eeq

The SAC algorithm uses a stochastic policy, i.e., the actor gives the conditional probability of an action given the state. In order to train the actor, we sample from the posterior probability $\pi_{\phi}(a|s)$ and $\log \pi_{\phi}(a|s)$ is modeled as a mapping $f(\epsilon,s)$, with reparametrization noise $\epsilon$ \citep{VAE}. We perform gradient descent with the gradient
\beqn \label{gradASAC}
\lefteqn{\nabla_{\phi} J(\phi)=\nabla_\phi \log \pi_{\phi}(a|s)}\\\nonumber
&&+ \nabla_a ( \log \pi_{\phi}(a|s) - Q(s,a))|_{a=f(\epsilon,s)} \times \nabla_\phi f(\epsilon,s).
\eeqn

In practice, RL is prone to instability and divergence, and therefore both TD3 and SAC use some of the following refinements \citep{DQN,DDPG,TD3}:
\begin{itemize}
\item Instead of having one critic, TD3 has two online  ($Q_{\xi_1}(s,a)$ and $Q_{\xi_2}(s,a)$) and two target ($Q_{\overline{\xi}_1}(s^\prime,a^\prime)$ and $Q_{\overline{\xi}_2}(s^\prime,a^\prime)$) critics and selects the lower evaluation out of the two to evaluate (\ref{bellman}). Similarly, SAC uses two online critics.
\item TD3 uses both an online and a target actor. When evaluating (\ref{bellman}), we use the target network to evaluate $\underset{a^\prime=\pi(s^\prime)}{\mathrm max} Q(s^\prime,a^\prime)$.
\item SAC uses both an online and a target value function. When evaluating (\ref{bellman}), $\underset{a^\prime=\pi(s^\prime)}{\mathrm max} Q(s^\prime,a^\prime)$ is replaced with  $V_{\overline{\zeta}}(s)$.
\item The parameters of the target networks are updated by a weighted average between the online and target parameters, and for the target actor in TD3, the update is only performed at delayed intervals.
\item TD3 uses random actions at the start of learning for exploration, which is called the warmup period.
\item Both TD3 and SAC keep a buffer to store past transitions: $(s,a,r,s^\prime)$ that we re-use in mini-batch mode sampling during training (this is the replay buffer in Algorithm \ref{algRL}).
\end{itemize}

We refer the reader to \cite{TD3} and \cite{SAC} for detailed descriptions of TD3 and SAC. In section \ref{sec:simul}, we compare their performance in training our RL agent.

\section{Simulations}\label{sec:simul}
{ As seen in Algorithm \ref{algRL}, we need simulations to train the RL agent. In contrast to conventional search-based methods in hyperparamter tuning (e.g., \cite{bergstra12a,Bates2021}), the training phase of the RL agent needs additional computations. In a pipeline setting (Fig. \ref{agent_pipeline}), the tuning of hyperparemters needs to be done as fast as possible so that the flow of the data is not delayed. Therefore, after the training phase, our RL agent should run faster than conventional search-based methods. In section \ref{ssec:simul_enet}, to compare the performance of the RL agent with the performance of traditional grid search-based hyperparameter optimization, we consider an analogous but simpler pipeline than the calibration pipeline. We also compare the learning rates of DDPG, TD3 and SAC. Thereafter, in section \ref{ssec:simul_calib}, we describe the simulation setup to train the RL agent in calibration pipelines. We show the increase in reward (or score) of the RL agent as we perform more simulations, hence confirming that the RL agent is able to learn. }

\subsection{Elastic net regression}\label{ssec:simul_enet}
In elastic net regression \citep{ElasticNet}, we observe ${\bf x}$ ($\in \mathbb{R}^M$) given the design matrix ${\bf A}$ ($\in \mathbb{R}^{M\times M}$) and estimate the parameters ${\bmath \theta}$ ($\in \mathbb{R}^M$) of the linear model ${\bf x}={\bf A}{\bmath \theta}$. The estimation is performed under regularization as
\beq \label{enet}
\widehat{{\bmath \theta}}=\underset{\bmath \theta}{\argmin}\left(\|{\bf x}-{\bf A}{\bmath \theta}\|^2 + \rho_2 \|{\bmath \theta}\|^2+ \rho_1 \|{\bmath \theta}\|_1 \right)
\eeq
where we have hyperparameters $\rho_1$ and $\rho_2$. We can easily relate (\ref{enet}) to (\ref{aug}), which is our original problem. Moreover, (\ref{enet}) can also be considered as a simplified image deconvolution problem (see e.g., \cite{Ammanouil2019}), where ${\bf x}$ is the visibility data, ${\bmath \theta}$ is the image and ${\bf A}$ is the de-gridding and Fourier transform operations (note that ${\bf A}$ will not be a square matrix in this case).

Representing (\ref{enet}) as in (\ref{smodel}) and using (\ref{calA}), for the elastic net problem, we have
\beq \label{calAA}
\mathcal{A} = {\bf A}\frac{1}{2}\left({\bf A}^T{\bf A}+(\rho_2+\rho_1 \delta(\|{\bmath \theta}\|)) {\bf I}\right)^{-1}\left(-2{\bf A}^T\right)
\eeq
where $\delta(\cdot)$ is the Dirac delta function. We can use (\ref{calAA}) to calculate the influence function. The RL agent for the elastic net regression hyperparameter selection interacts with its environment as follows:
\begin{itemize}
\item State: design matrix ${\bf A}$ and eigenvalues $(1+\lambda(\mathcal{A}))$.
\item Action: regularization factors $\rho_1,\rho_2$.
\item Reward: $\frac{\|{\bf x}\|}{\|{\bf x}-{\bf A}{\bmath \theta}\|}+\frac{min(1+\lambda(\mathcal{A}))}{max(1+\lambda(\mathcal{A}))}$.
\end{itemize} 
Note that the reward considers both the (inverse) residual error as well as the bias (ratio of eigenvalues of the influence matrix). In this manner, we can achieve a balance between the bias and the variance of the residual.

For the simulations, we consider $M=20$ and the agent is implemented in Pytorch \citep{paszke}, within an {\em openai.gym} compatible environment \footnote{https://gym.openai.com/}. We use the limited-memory Broyden Fletcher Goldfarb Shanno (LBFGS) algorithm \cite{DSW2019} to solve (\ref{enet}) and the Adam optimizer \cite{Adam} is used to train the agent.

The networks used for TD3 (and DDPG) RL agent are as follows. The critic uses two linear layers each for the state and for the action, finally combined in one linear layer to produce a scalar output. The actor uses four linear layers to produce the two dimensional output. We use exponential linear unit activation \citep{ELU} except in the last layer of the actor, where we use hyperbolic tangent activation (which is scaled to the valid range afterwards). We use batch normalization \citep{batchnorm} between all layers except the last. In the SAC RL agent, the actor produces two outputs, the mean and the standard deviation for the conditional probability of the action; apart from this the SAC actor is similar to the TD3 actor. The critic network is similar to the one in TD3. The value network is also similar to the critic, except that it only takes the state as input (no action is needed as input). 

In each episode $e$ in Algorithm \ref{algRL}, we create ${\bf A}$ by filling its entries with unit variance, zero mean Gaussian random values. The ground truth value of the parameters, i.e. ${\bmath \theta}_0$, is generated with the number of non-zero entries randomly generated in the range $[3,M]$ and these non-zero entries are filled with unit variance, zero mean Gaussian random values. In order to get ${\bf x}$, we add unit variance, zero mean Gaussian noise to ${\bf A}{\bmath \theta}_0$ with a signal to noise ratio of $0.1$.

\begin{figure}
\begin{minipage}{0.98\linewidth}
\begin{center}
\centering
\centerline{\epsfig{figure=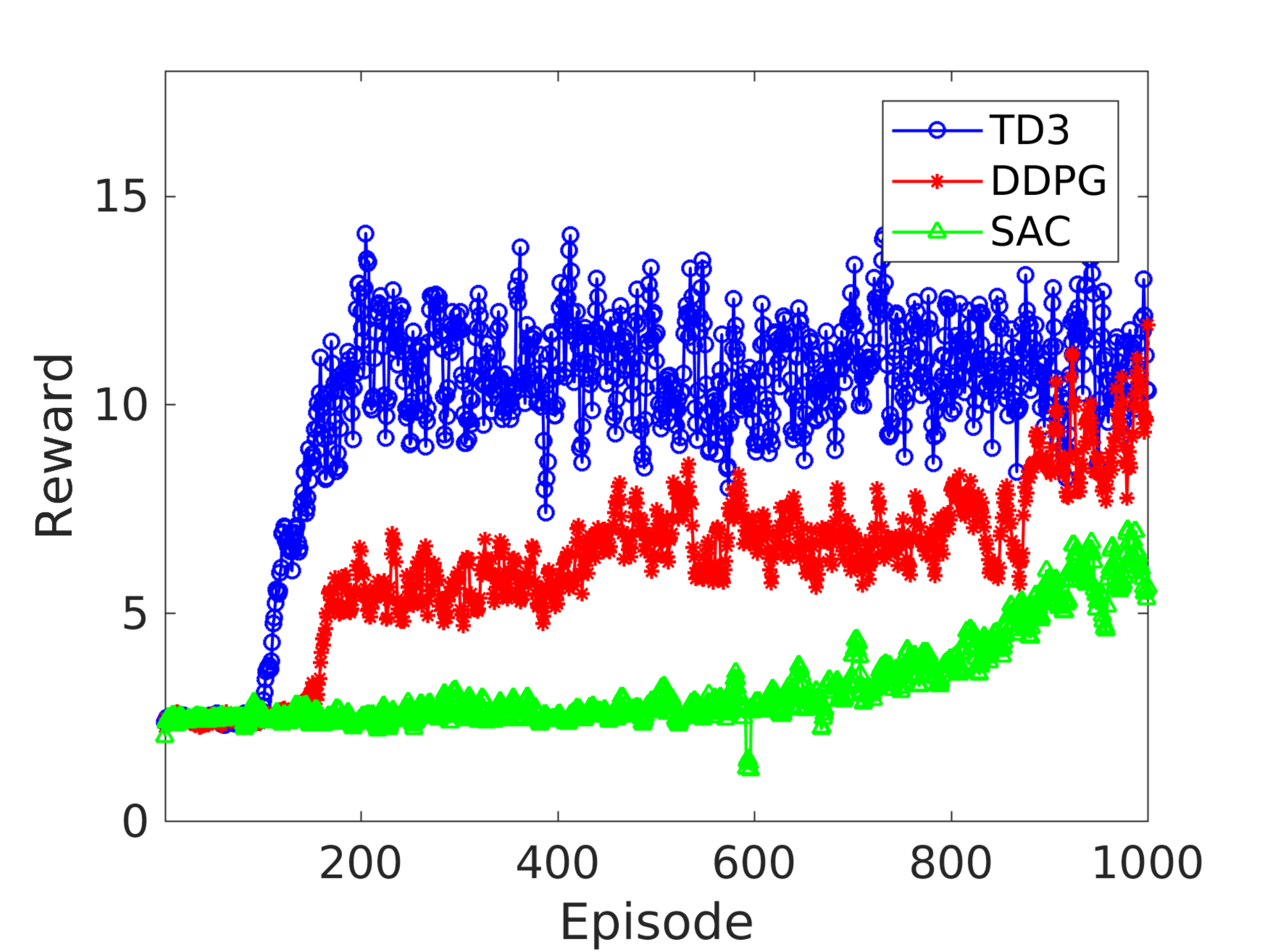,width=8.0cm}}
\vspace{0.1in}
(a) $L=1$
\end{center}
\end{minipage}\\
\begin{minipage}{0.98\linewidth}
\begin{center}
\centering
\centerline{\epsfig{figure=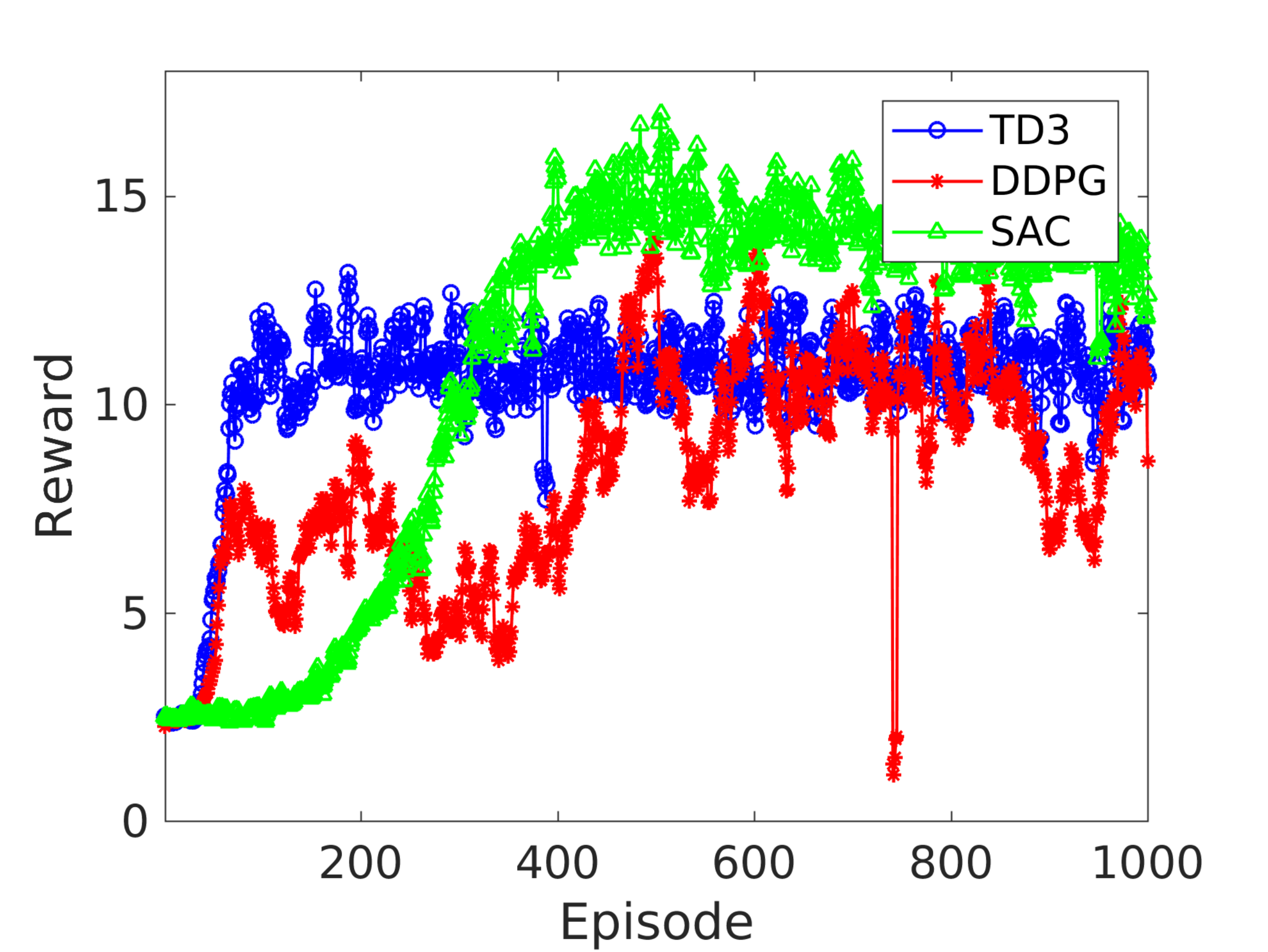,width=8.0cm}}
\vspace{0.1in}
(b) $L=4$
\end{center}
\end{minipage}
\begin{minipage}{0.98\linewidth}
\begin{center}
\centering
\centerline{\epsfig{figure=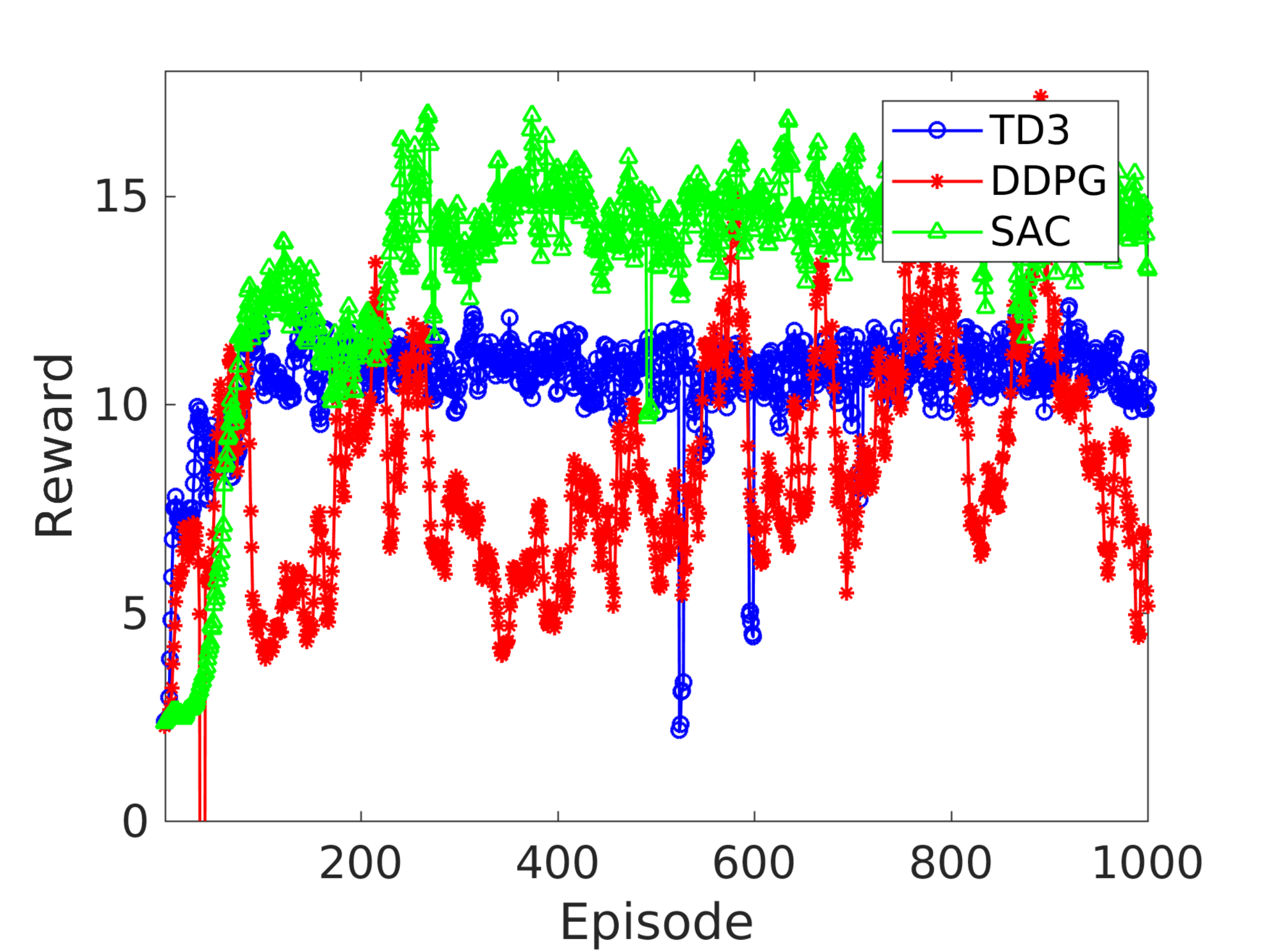,width=8.0cm}}
\vspace{0.1in}
(c) $L=20$
\end{center}
\end{minipage}
\caption{The score variation with episode for the agent selecting hyperparameters in elastic net regression. The number of iterations for each episode $L$ is varied from (a) $L=1$ (b) $L=4$ and (c) $L=20$. We see that TD3 algorithm learns faster, but SAC reaches a higher reward for large number of episodes.\label{elastic_net}}
\end{figure}

For each episode $e$, we limit the number of iterations to $L$. For TD3, we use a warmup interval of 100 iterations. In Fig. \ref{elastic_net}, we show the reward after each episode for different values of $L$. We have compared DDPG, TD3 and SAC algorithms in Fig. \ref{elastic_net}. We clearly see that both TD3 and SAC display stable learning after a certain number of episodes, unlike DDPG. We also see that SAC reaches a higher reward level, however at a lower rate than TD3.

\begin{figure}
\begin{minipage}{0.98\linewidth}
\begin{center}
\centering
\centerline{\epsfig{figure=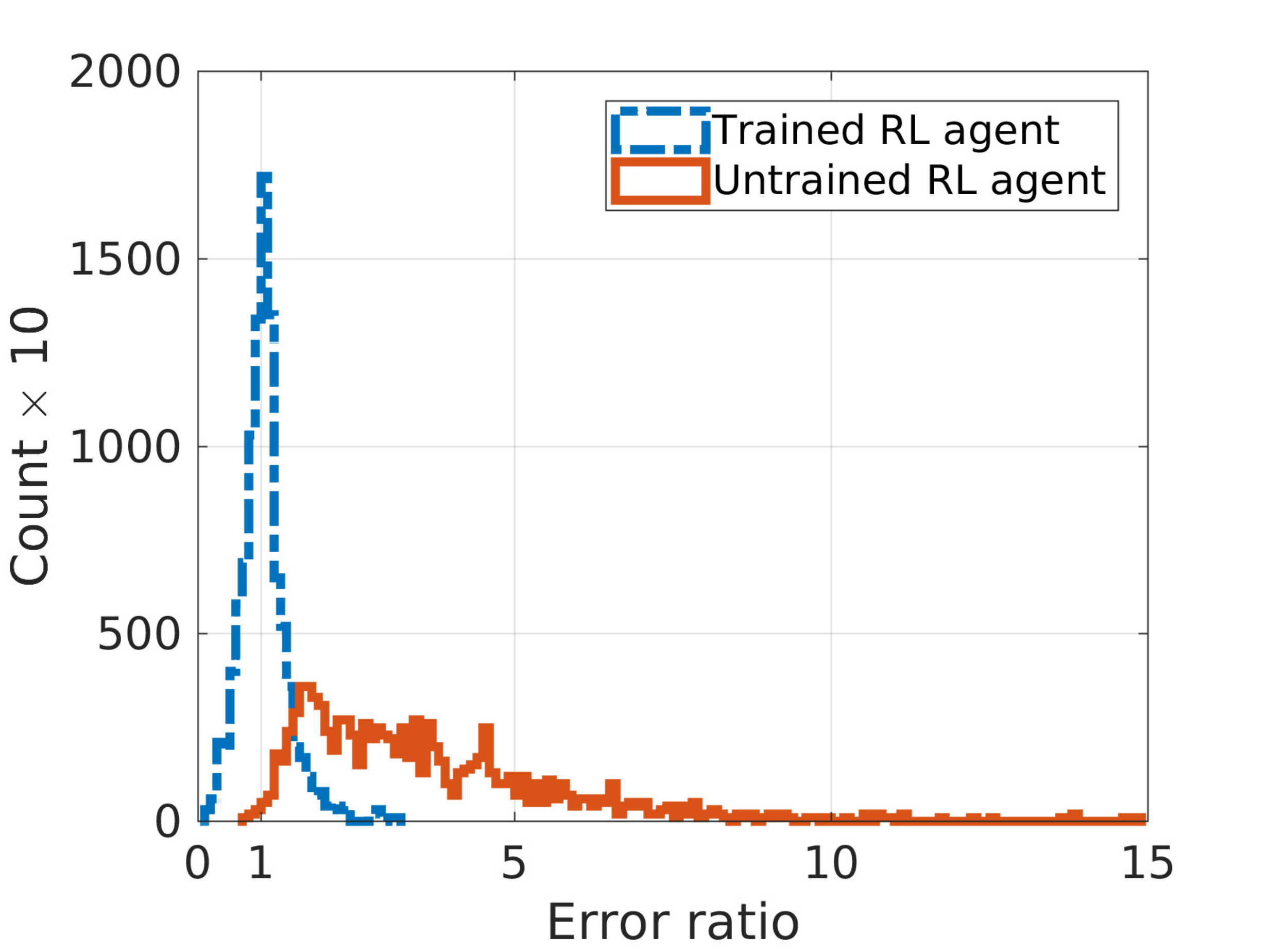,width=8.0cm}}
\end{center}
\end{minipage}\\
\caption{Histogram comparison of the error ratio for hyperparameter tuning using grid search against using TD3 RL agent. Results using both the untrained and trained RL agent are shown. With the trained RL agent, we get a mean result of about $1$ indicating that the RL agent is able to predict as equally well as using grid search.\label{elastic_comparison}}
\end{figure}

As shown in Fig. \ref{elastic_net}, SAC RL agent reaches the highest reward level but TD3 achieves a stable reward earlier. We consider the TD3 RL agent, trained with $L=4$ and $E=1000$ using Algorithm \ref{algRL}, and compare its performance with traditional grid search based hyperparamter tuning. For the grid search, we use a grid of values $[0.001, 0.005, 0.01, 0.05, 0.1]$ for $\rho_1$ and $\rho_2$. For the RL agent, we use $L=4$ iterations to make a prediction for $\rho_1$ and $\rho_2$ while for the grid search we need at least $5\times 5=25$ evaluations of (\ref{enet}). We compare the error ratio $||{\bmath \theta}_{RL}-{\bmath \theta}_0||/||{\bmath \theta}_{GR}-{\bmath \theta}_0||$ where ${\bmath \theta}_0$ is the ground truth, ${\bmath \theta}_{RL}$ is the solution obtained with hyperparameters determined by the RL agent and ${\bmath \theta}_{GR}$ is the solution obtained with hyperparameters determined by grid search. We show histograms of this error ratio for $1000$ episodes in Fig. \ref{elastic_comparison}.

As seen in Fig. \ref{elastic_comparison}, the trained RL agent is able to perform equally well as grid search, however the RL agent needs only $4$ evaluation to make this prediction while grid search uses $25$ evaluations. This makes a strong case for using RL for hyperparameter tuning, especially in a situation where data is processed in a pipeline. 

\subsection{Calibration}\label{ssec:simul_calib}
We give a summary of the calibration pipeline, i.e. the environment in Fig. \ref{agent_pipeline}. { The observation consists of $N=62$ stations.} The data passed through the calibration pipeline have $1$ sec integration time and $8$ frequencies in the range $[115,185]$ MHz. Calibration is performed for every $\Delta_t=10$ time samples (every $10$ sec, $D=151\ 280$) and $6$ such calibrations are performed to interact with the RL agent. In other words, $1$ min of data are used by the agent to make a recommendation on the regularization parameters. In practice, the first $1$ min of data is used to adjust the hyperparameters of the pipeline, and thereafter, the pipeline is run unchanged for the full observation (which typically will last a few hours). We use SAGECal\footnote{http://sagecal.sourceforge.net/} for calibration and excon\footnote{https://sourceforge.net/projects/exconimager/} for imaging.

We simulate data for calibration along $K=4$ directions in the sky, with systematic errors ${\bf J}_{kpf}$ in (\ref{V}) that are smooth both in time and in frequency, but have random initial values. One direction out of $K$ represents the observed field, and its location is randomly chosen within $1$ deg radius from the field centre. The remaining $K-1$ directions play the role of outlier sources (for instance sources such as the Sun, Cassiopeia A, Cygnus A) and are randomly located within a $15$ deg radius from the field centre. The intensity of the central source is randomly selected within $[3,10]$ Jy. The intrinsic intensities of the outlier sources are randomly selected within $[100,250]$ Jy. An additional $200$ weak sources (point sources and Gaussians) with intensities in $[0.01,0.1]$ Jy { ($N(S) \propto S^{-2}$, $N(S)$ being the source count and $S$ being the flux density)} are randomly positioned across a $16\times 16$ square degrees surrounding the field centre. { The spectral indices of the $K$ sources being calibrated are generated from a standard normal distribution and all the weak sources have flat spectra.} The simulation includes the effect due to the station beam shape. Finally, zero mean, complex circular { symmetric} Gaussian noise is added to the simulated data with a signal to noise ratio of $0.05$. Note that the sky model used for calibration only includes the $K$ sources (not the weak sources) and the $K-1$ outlier sources have their intensities divided by a factor of $100$ to account for the receiver beam attenuation (apparent value).

The RL agent used in the calibration pipeline interacts with the environment (or the pipeline) as follows:
\begin{itemize}
\item State: the sky model used in calibration and the influence map $\mathcal{I}({\bmath \theta})$ generated using $1$ min of data.
\item Action: regularization factors $\rho_k$ for $K$ directions.
\item Reward: $\frac{\sigma_0}{\sigma_1}+\frac{1}{\sigma(\mathcal{I}({\bmath \theta}))}$, where $\sigma_0$ and $\sigma_1$ are the image noise standard deviations of the input data and output residual, respectively. { The standard deviation of $\mathcal{I}({\bmath \theta})$ is given by $\sigma(\mathcal{I}({\bmath \theta}))$.}
\end{itemize}

\begin{figure}
\begin{minipage}{0.98\linewidth}
\begin{center}
\centering
\centerline{\epsfig{figure=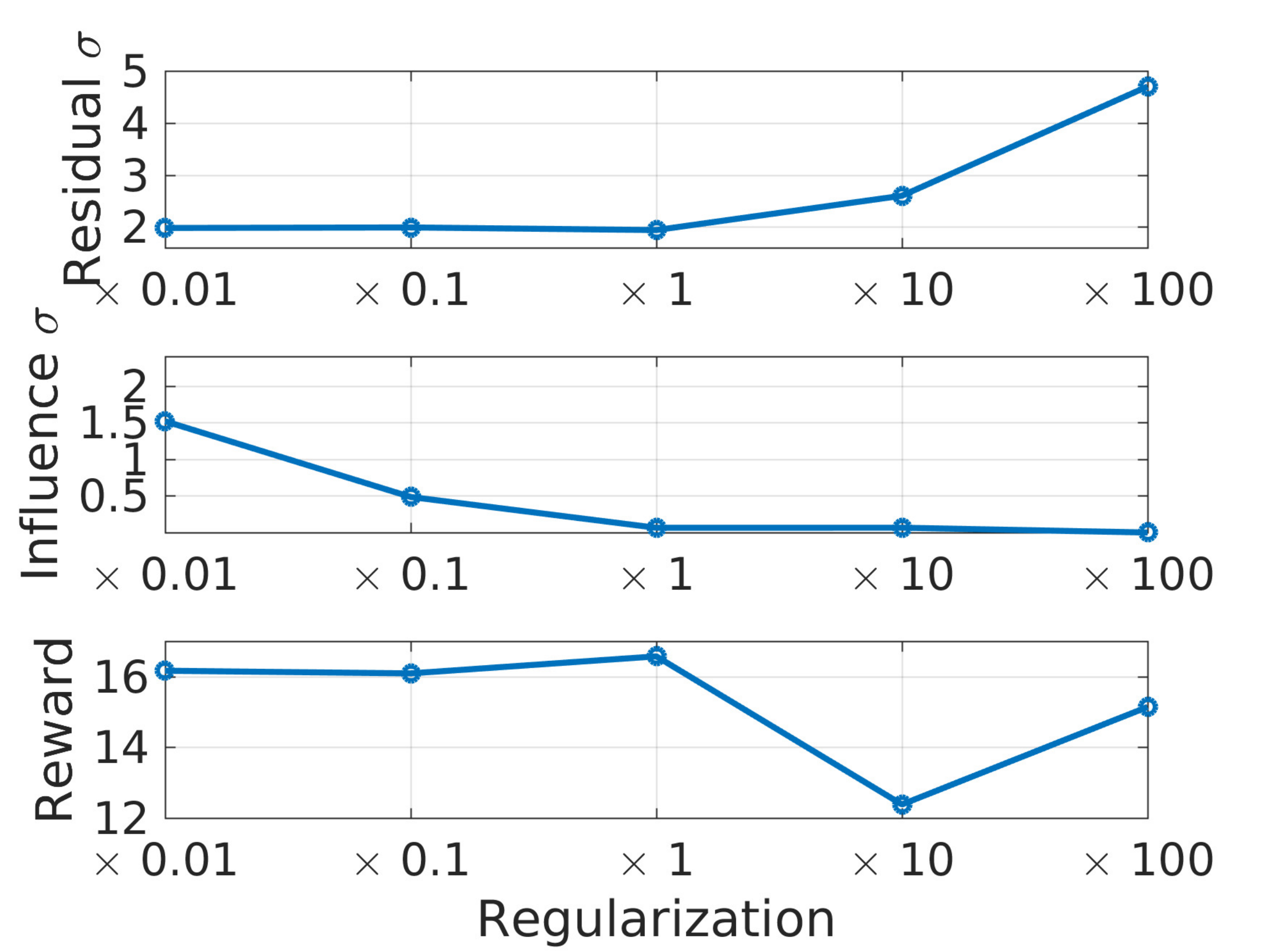,width=8.1cm}}
\vspace{0.1in}
\end{center}
\end{minipage}
\caption{The residual noise, the influence map noise and the reward for various values of regularization. The regularization factors are scaled up and down from the optimal values ($\times 1$).\label{reward}}
\end{figure}

In Fig. \ref{reward}, we show the residual image standard deviation, influence map standard deviation and the calculated reward. The $x$-axis corresponds to the regularization parameters $\rho_k$, scaled up and down by factors of ten (starting from the regularization that gives the highest reward).
For low regularization values, we see more or less constant residual image noise, however, we clearly see a high level of influence. This high influence on the left hand side of Fig. \ref{reward} implies high variance of $\widehat{{\bmath \theta}}$ (due to the unmodeled sources in the sky) and as a consequence, we should expect high level of weak signal suppression. On the other hand, on the right hand side of Fig. \ref{reward}, for high regularization, we see increased residual image noise, meaning poor calibration. We compose our reward to reach a balance between the bias and variance of the residual.

We use both TD3 and SAC for training our RL agent. The RL agents and the environment are quite similar to the ones used in elastic net regression. In TD3, the critic uses 3 convolutional layers to process the influence map, and two linear layers to process the action and the sky model. These two are combined at the final linear layer. The actor also uses 3 convolutional layers to process the influence map and two linear layers for the sky model. Once again, they are combined at the final linear layer. The first three layers of both the actor and critic use batch normalization. The activation functions are exponential linear units except at the last layer of the actor, where we use hyperbolic tangent activation. The action is scaled to the required range by the environment. In the SAC RL agent, the critic is similar to the one in the TD3 RL agent. The actor is almost the same, except it produces two outputs, for the mean and the standard distribution of the conditional probability of the action. The value network is similar to the critic, except it does not process the action as an input.

\begin{figure}
\begin{minipage}{0.98\linewidth}
\begin{center}
\centering
\centerline{\epsfig{figure=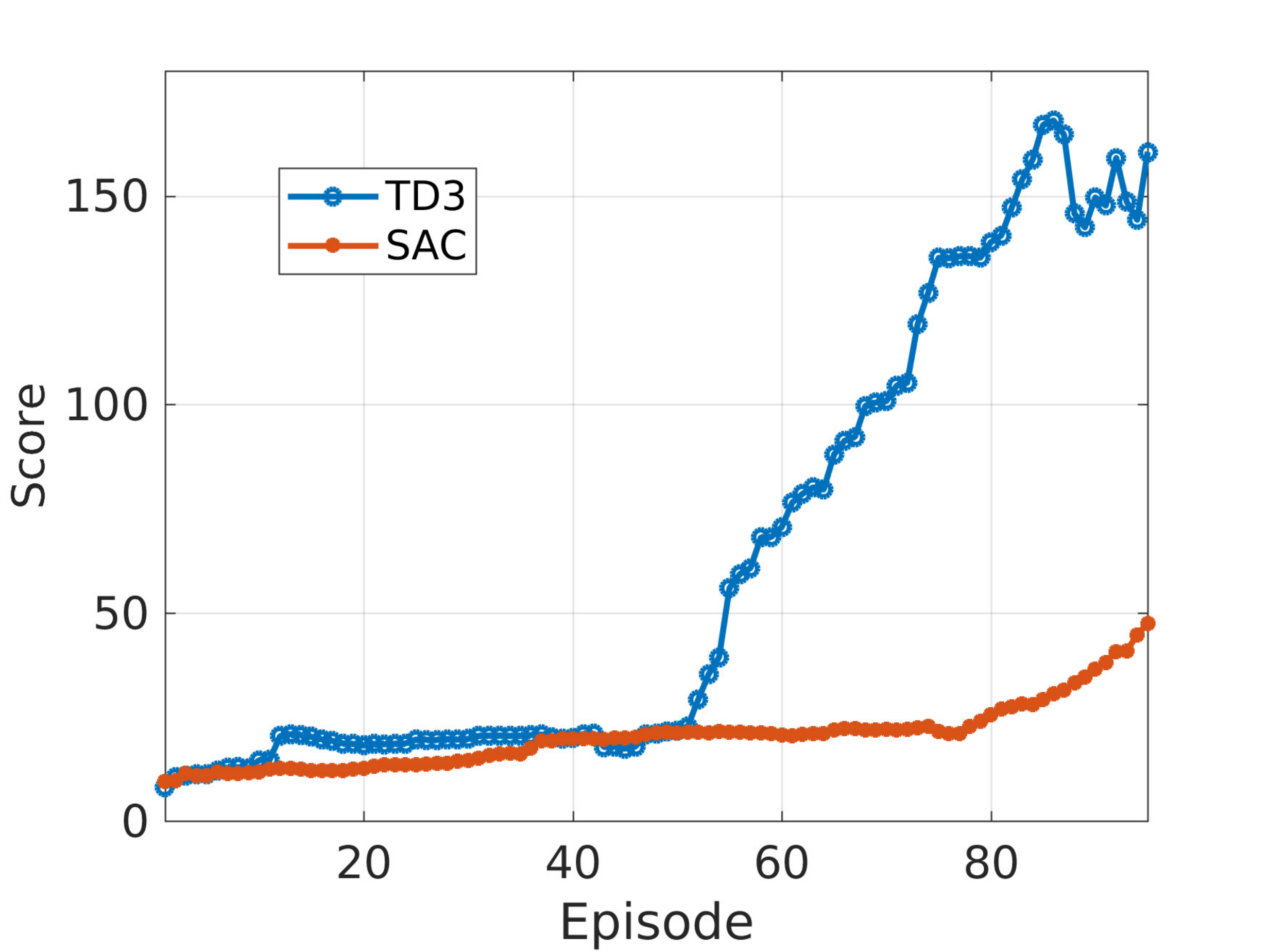,width=8.1cm}}
\vspace{0.1in}
\end{center}
\end{minipage}
\caption{The variation of the score (or the reward) with episode for the TD3 and SAC RL agents in calibration. The number of iterations for each episode is $L=4$. The TD3 RL agent is able to reach a higher score much faster than the SAC RL agent.\label{scores}}
\end{figure}

The training of our RL agents are done as in Algorithm \ref{algRL}. In each episode $e$, we generate a random sky model and data as described previously. The pipeline is run with initial regularization factors determined analytically \citep{EUSIPCO}. The state information generated by the pipeline is fed into the RL agent to retrieve updated regularization factors. The pipeline is run again and we repeat the cycle. Using the results shown in Fig. \ref{elastic_net}, we limit the number of such iterations per each episode to $L=4$. At the final iteration, we record the reward (or the score) to measure the progress of the agent. The score variation over a number of such episodes is shown in Fig. \ref{scores}. 

After about $50$ episodes, we see the TD3 RL agent reaching a higher score (the average reward at each episode) in Fig. \ref{scores}, indicating that the agent is able to learn and improve its recommendations for the regularization factors to use. The SAC RL agent is learning at a lower rate, and this behavior is similar to the elastic net regression RL agent, as seen in Fig. \ref{elastic_net} (b). Once we have a fully trained agent, following Algorithm \ref{algRL}, we omit the simulation step (line 4). The state is determined using $1$ min of data and we use $L=4$ iterations to make a recommendation for the hyperparameters. We will get a result similar to Fig. \ref{influence_maps} bottom row with the recommendation. We can keep the pipeline running with the same hyperparameters for a much longer time or dynamically adjust the hyperparamters of the pipeline by interacting with the RL agent. This aspect needs more work with real data, which we intend to pursue as future work. In addition, we can develop similar agents for other data processing steps in radio astronomy, including RFI mitigation and image synthesis. Extending beyond radio astronomy, similar techniques can be applied in other data processing pipelines in general machine learning problems.

\section{Conclusions}\label{sec:conc}
We have introduced the use of reinforcement learning in hyperparemeter tuning of radio astronomical calibration pipelines. By using both the input-output noise reduction as well as the influence function to determine the reward offered to the RL agent, we can reach a balance between the bias and the variance introduced by the pipeline. As illustrated by the elastic-net regression example, we can also apply the same technique to many other data processing steps, not only in radio astronomy but also in other applications. 

\section*{Acknowledgments}
We thank the anonymous reviewer for the careful review and valuable comments.

\section*{Data availability}
Ready to use software based on this work and test data are available online\footnote{https://github.com/SarodYatawatta/smart-calibration}.

\bibliographystyle{mnras}
\bibliography{references}

\begin{thebibliography}{}
\makeatletter
\relax
\def\mn@urlcharsother{\let\do\@makeother \do\$\do\&\do\#\do\^\do\_\do\%\do\~}
\def\mn@doi{\begingroup\mn@urlcharsother \@ifnextchar [ {\mn@doi@}
  {\mn@doi@[]}}
\def\mn@doi@[#1]#2{\def\@tempa{#1}\ifx\@tempa\@empty \href
  {http://dx.doi.org/#2} {doi:#2}\else \href {http://dx.doi.org/#2} {#1}\fi
  \endgroup}
\def\mn@eprint#1#2{\mn@eprint@#1:#2::\@nil}
\def\mn@eprint@arXiv#1{\href {http://arxiv.org/abs/#1} {{\tt arXiv:#1}}}
\def\mn@eprint@dblp#1{\href {http://dblp.uni-trier.de/rec/bibtex/#1.xml}
  {dblp:#1}}
\def\mn@eprint@#1:#2:#3:#4\@nil{\def\@tempa {#1}\def\@tempb {#2}\def\@tempc
  {#3}\ifx \@tempc \@empty \let \@tempc \@tempb \let \@tempb \@tempa \fi \ifx
  \@tempb \@empty \def\@tempb {arXiv}\fi \@ifundefined
  {mn@eprint@\@tempb}{\@tempb:\@tempc}{\expandafter \expandafter \csname
  mn@eprint@\@tempb\endcsname \expandafter{\@tempc}}}

\bibitem[\protect\citeauthoryear{Ammanouil, Ferrari, Mary, Ferrari  \&
  Loi}{Ammanouil et~al.}{2019}]{Ammanouil2019}
Ammanouil R.,  Ferrari A.,  Mary D.,  Ferrari C.,   Loi F.,  2019, Monthly
  Notices of the Royal Astronomical Society, 490, 37

\bibitem[\protect\citeauthoryear{{Bates}, {Hastie}  \& {Tibshirani}}{{Bates}
  et~al.}{2021}]{Bates2021}
{Bates} S.,  {Hastie} T.,   {Tibshirani} R.,  2021, arXiv e-prints, \href
  {https://ui.adsabs.harvard.edu/abs/2021arXiv210400673B} {p. arXiv:2104.00673}

\bibitem[\protect\citeauthoryear{Bergstra \& Bengio}{Bergstra \&
  Bengio}{2012}]{bergstra12a}
Bergstra J.,  Bengio Y.,  2012, Journal of Machine Learning Research, 13, 281

\bibitem[\protect\citeauthoryear{Boyd, Parikh, Chu, Peleato  \& Eckstein}{Boyd
  et~al.}{2011}]{boyd2011}
Boyd S.,  Parikh N.,  Chu E.,  Peleato B.,   Eckstein J.,  2011, Foundations
  and Trends{\textregistered} in Machine Learning, 3, 1

\bibitem[\protect\citeauthoryear{{Brossard}, {El Korso}, {Pesavento}, {Boyer},
  {Larzabal}  \& {Wijnholds}}{{Brossard} et~al.}{2016}]{Brossard2016}
{Brossard} M.,  {El Korso} M.~N.,  {Pesavento} M.,  {Boyer} R.,  {Larzabal} P.,
    {Wijnholds} S.~J.,  2016, preprint, \href
  {http://adsabs.harvard.edu/abs/2016arXiv160902448B} {} (\mn@eprint {arXiv}
  {1609.02448})

\bibitem[\protect\citeauthoryear{{Clevert}, {Unterthiner}  \&
  {Hochreiter}}{{Clevert} et~al.}{2015}]{ELU}
{Clevert} D.-A.,  {Unterthiner} T.,   {Hochreiter} S.,  2015, arXiv e-prints,
  \href {http://adsabs.harvard.edu/abs/2015arXiv151107289C} {}

\bibitem[\protect\citeauthoryear{Cook \& Weisberg}{Cook \&
  Weisberg}{1982}]{cook1982residuals}
Cook R.,  Weisberg S.,  1982, Residuals and Influence in Regression.
Monographs on statistics and applied probability, Chapman \& Hall, \url
  {http://books.google.nl/books?id=MVSqAAAAIAAJ}

\bibitem[\protect\citeauthoryear{{Fujimoto}, {van Hoof}  \& {Meger}}{{Fujimoto}
  et~al.}{2018}]{TD3}
{Fujimoto} S.,  {van Hoof} H.,   {Meger} D.,  2018, arXiv e-prints, \href
  {https://ui.adsabs.harvard.edu/abs/2018arXiv180209477F} {p. arXiv:1802.09477}

\bibitem[\protect\citeauthoryear{Geman, Bienenstock  \& Doursat}{Geman
  et~al.}{1992}]{neco1992}
Geman S.,  Bienenstock E.,   Doursat R.,  1992, \mn@doi [Neural Computation]
  {10.1162/neco.1992.4.1.1}, 4, 1

\bibitem[\protect\citeauthoryear{{Haarnoja}, {Zhou}, {Abbeel}  \&
  {Levine}}{{Haarnoja} et~al.}{2018}]{SAC}
{Haarnoja} T.,  {Zhou} A.,  {Abbeel} P.,   {Levine} S.,  2018, arXiv e-prints,
  \href {https://ui.adsabs.harvard.edu/abs/2018arXiv180101290H} {p.
  arXiv:1801.01290}

\bibitem[\protect\citeauthoryear{{Hamaker}, {Bregman}  \& {Sault}}{{Hamaker}
  et~al.}{1996}]{HBS}
{Hamaker} J.~P.,  {Bregman} J.~D.,   {Sault} R.~J.,  1996, Astronomy and
  Astrophysics Supp., 117, 137

\bibitem[\protect\citeauthoryear{Hampel, Ronchetti, Rousseeuw  \&
  Stahel}{Hampel et~al.}{1986}]{Hampel86}
Hampel F.~R.,  Ronchetti E.,  Rousseeuw P.~J.,   Stahel W.~A.,  1986, Robust
  statistics: the approach based on influence functions.
New York USA:Wiley, \url {https://archive-ouverte.unige.ch/unige:23238}

\bibitem[\protect\citeauthoryear{Hestenes}{Hestenes}{1969}]{Hestenes69}
Hestenes M.~R.,  1969, Journal of Optimization Theory and Applications, 4, 303

\bibitem[\protect\citeauthoryear{{Ioffe} \& {Szegedy}}{{Ioffe} \&
  {Szegedy}}{2015}]{batchnorm}
{Ioffe} S.,  {Szegedy} C.,  2015, arXiv e-prints, \href
  {https://ui.adsabs.harvard.edu/abs/2015arXiv150203167I} {p. arXiv:1502.03167}

\bibitem[\protect\citeauthoryear{{Kingma} \& {Ba}}{{Kingma} \&
  {Ba}}{2014}]{Adam}
{Kingma} D.~P.,  {Ba} J.,  2014, preprint, \href
  {http://adsabs.harvard.edu/abs/2014arXiv1412.6980K} {} (\mn@eprint {arXiv}
  {1412.6980})

\bibitem[\protect\citeauthoryear{{Kingma} \& {Welling}}{{Kingma} \&
  {Welling}}{2013}]{VAE}
{Kingma} D.~P.,  {Welling} M.,  2013, arXiv e-prints, \href
  {https://ui.adsabs.harvard.edu/abs/2013arXiv1312.6114K} {p. arXiv:1312.6114}

\bibitem[\protect\citeauthoryear{Koh \& Liang}{Koh \& Liang}{2017}]{Koh17}
Koh P.~W.,  Liang P.,  2017, in Precup D.,  Teh Y.~W.,  eds,  Proceedings of
  Machine Learning Research Vol. 70, Proceedings of the 34th International
  Conference on Machine Learning. PMLR, International Convention Centre,
  Sydney, Australia, pp 1885--1894, \url
  {http://proceedings.mlr.press/v70/koh17a.html}

\bibitem[\protect\citeauthoryear{LeCun, Bengio  \& Hinton}{LeCun
  et~al.}{2015}]{Lecun2015}
LeCun Y.,  Bengio Y.,   Hinton G.,  2015, Nature, 521, 436 EP

\bibitem[\protect\citeauthoryear{{Lillicrap}, {Hunt}, {Pritzel}, {Heess},
  {Erez}, {Tassa}, {Silver}  \& {Wierstra}}{{Lillicrap} et~al.}{2015}]{DDPG}
{Lillicrap} T.~P.,  {Hunt} J.~J.,  {Pritzel} A.,  {Heess} N.,  {Erez} T.,
  {Tassa} Y.,  {Silver} D.,   {Wierstra} D.,  2015, arXiv e-prints, \href
  {https://ui.adsabs.harvard.edu/abs/2015arXiv150902971L} {p. arXiv:1509.02971}

\bibitem[\protect\citeauthoryear{{McMahan}, {Moore}, {Ramage}, {Hampson}  \&
  {Ag{\"u}era y Arcas}}{{McMahan} et~al.}{2016}]{McMahan2016}
{McMahan} B.~H.,  {Moore} E.,  {Ramage} D.,  {Hampson} S.,   {Ag{\"u}era y
  Arcas} B.,  2016, arXiv e-prints, \href
  {https://ui.adsabs.harvard.edu/abs/2016arXiv160205629B} {p. arXiv:1602.05629}

\bibitem[\protect\citeauthoryear{Mertens et~al.,}{Mertens
  et~al.}{2020}]{mertens2020}
Mertens F.,  et~al., 2020, Monthly Notices of the Royal Astronomical Society,
  493, 1662

\bibitem[\protect\citeauthoryear{Mevius et~al.}{Mevius
  et~al.}{2021}]{mevius2021}
Mevius M.,  et~al., 2021, pre-print

\bibitem[\protect\citeauthoryear{Mnih et~al.,}{Mnih et~al.}{2015}]{Atari}
Mnih V.,  et~al., 2015, Nature, 518, 529

\bibitem[\protect\citeauthoryear{Ollier, Korso, Ferrari, Boyer  \&
  Larzabal}{Ollier et~al.}{2018}]{OLLIER2018}
Ollier V.,  Korso M. N.~E.,  Ferrari A.,  Boyer R.,   Larzabal P.,  2018,
  \mn@doi [Signal Processing] {https://doi.org/10.1016/j.sigpro.2018.07.024},
  153, 348

\bibitem[\protect\citeauthoryear{Paszke et~al.,}{Paszke et~al.}{2017}]{paszke}
Paszke A.,  et~al., 2017, in NIPS-W.

\bibitem[\protect\citeauthoryear{{Patil} et~al.,}{{Patil}
  et~al.}{2017}]{Patil2017}
{Patil} A.~H.,  et~al., 2017, \mn@doi [\apj] {10.3847/1538-4357/aa63e7}, 838,
  65

\bibitem[\protect\citeauthoryear{Powell}{Powell}{1969}]{Powell69}
Powell M.,  1969, Optimization, pp 283--298

\bibitem[\protect\citeauthoryear{Silver, Lever, Heess, Degris, Wierstra  \&
  Riedmiller}{Silver et~al.}{2014}]{Silver2014}
Silver D.,  Lever G.,  Heess N.,  Degris T.,  Wierstra D.,   Riedmiller M.,
  2014, in Proceedings of the 31st International Conference on International
  Conference on Machine Learning - Volume 32. ICML 2014.
JMLR.org, pp 387--395

\bibitem[\protect\citeauthoryear{Sutton \& Barto}{Sutton \&
  Barto}{2018}]{SuttonBarto}
Sutton R.~S.,  Barto A.~G.,  2018, Reinforcement Learning: An Introduction.
A Bradford Book, Cambridge, MA, USA

\bibitem[\protect\citeauthoryear{{Thompson}, {Moran}  \& {Swenson}}{{Thompson}
  et~al.}{2001}]{TMS}
{Thompson} A.,  {Moran} J.,   {Swenson} G.,  2001, {Interferometry and
  synthesis in radio astronomy (3rd ed.)}.
Wiley Interscience, New York

\bibitem[\protect\citeauthoryear{{Yatawatta}}{{Yatawatta}}{2015}]{DCAL}
{Yatawatta} S.,  2015, \mn@doi [\mnras] {10.1093/mnras/stv596}, 449, 4506

\bibitem[\protect\citeauthoryear{Yatawatta}{Yatawatta}{2016}]{EUSIPCO}
Yatawatta S.,  2016, in 24th European Signal Processing Conference (EUSIPCO),
  2016.

\bibitem[\protect\citeauthoryear{Yatawatta}{Yatawatta}{2018}]{SAM2018}
Yatawatta S.,  2018, in 2018 IEEE 10th Sensor Array and Multichannel Signal
  Processing Workshop (SAM). pp 485--489, \mn@doi{10.1109/SAM.2018.8448481}

\bibitem[\protect\citeauthoryear{Yatawatta}{Yatawatta}{2019}]{ST2019}
Yatawatta S.,  2019, \mn@doi [Monthly Notices of the Royal Astronomical
  Society] {10.1093/mnras/stz1222}, 486, 5646

\bibitem[\protect\citeauthoryear{Yatawatta}{Yatawatta}{2020}]{Y2020}
Yatawatta S.,  2020, \mn@doi [Monthly Notices of the Royal Astronomical
  Society] {10.1093/mnras/staa648}, 493, 6071

\bibitem[\protect\citeauthoryear{{Yatawatta}, {Diblen}, {Spreeuw}  \&
  {Koopmans}}{{Yatawatta} et~al.}{2018}]{DMUX}
{Yatawatta} S.,  {Diblen} F.,  {Spreeuw} H.,   {Koopmans} L.~V.~E.,  2018,
  \mn@doi [\mnras] {10.1093/mnras/stx3130}, \href
  {http://adsabs.harvard.edu/abs/2018MNRAS.475..708Y} {475, 708}

\bibitem[\protect\citeauthoryear{{Yatawatta}, {De Clercq}, {Spreeuw}  \&
  {Diblen}}{{Yatawatta} et~al.}{2019}]{DSW2019}
{Yatawatta} S.,  {De Clercq} L.,  {Spreeuw} H.,   {Diblen} F.,  2019, in 2019
  IEEE Data Science Workshop (DSW). pp 208--212,
  \mn@doi{10.1109/DSW.2019.8755567}

\bibitem[\protect\citeauthoryear{Zou \& Hastie}{Zou \&
  Hastie}{2005}]{ElasticNet}
Zou H.,  Hastie T.,  2005, \mn@doi [Journal of the Royal Statistical Society:
  Series B (Statistical Methodology)] {10.1111/j.1467-9868.2005.00503.x}, 67,
  301

\bibitem[\protect\citeauthoryear{{van Hasselt}, {Guez}  \& {Silver}}{{van
  Hasselt} et~al.}{2015}]{DQN}
{van Hasselt} H.,  {Guez} A.,   {Silver} D.,  2015, arXiv e-prints, \href
  {https://ui.adsabs.harvard.edu/abs/2015arXiv150906461V} {p. arXiv:1509.06461}

\makeatother
\end{thebibliography}
\bsp
\label{lastpage}
\end{document}